\documentclass[12pt]{article}
 
\usepackage{amsmath,amssymb} 
\usepackage{bm} %
\usepackage{cite} %
 
\usepackage{color} %
\usepackage{graphicx}

\usepackage[colorlinks]{hyperref} %
\hypersetup{
citecolor=blue,
linkcolor=blue,
urlcolor=blue
}

\usepackage{epstopdf}
\newcommand{\fig}[1]{Fig.~\ref{#1}}

\newcommand{\tab}[1]{Tab.~\ref{#1}}

\newcommand{\eq}[1]{Eq.~(\ref{#1})}
\newcommand{\eqs}[1]{Eqs.~(\ref{#1})}

\begin{document}
\title{
\vspace{-1.0truecm}
\bf Connecting the LHC  diphoton excess to \\
the Galatic center  gamma-ray excess 
}
\author{ 
Xian-Jun Huang,\ %
Wei-Hong Zhang,\ %
Yu-Feng Zhou \footnote{Email: huangxj@itp.ac.cn, whzhang@itp.ac.cn, yfzhou@itp.ac.cn} \\ \\
\textit{ Key Laboratory of Theoretical Physics,}\\
\textit{Institute of  Theoretical Physics,
Chinese Academy of Sciences}\\
\textit{ZhongGuanCun East Rd.55, Beijing, 100190, China}\\
\textit{School of Physical Sciences,University of Chinese Academy of Sciences}\\
\textit{ Yuquan Rd.19A, Beijing 100049, China}
}
\date{}
\maketitle %
\vspace{-1.0truecm}
\begin{abstract}
The recent LHC Run-2 data  have shown 
a possible excess in diphoton events, 
suggesting  the existence of  a new resonance $\phi$
with mass $M\sim 750$~GeV.  %
If $\phi$ plays the role of a portal particle connecting
the Standard Model  and the invisible dark sector,
the diphoton excess should  be correlated with another photon excess,
namely, 
the  excess in the diffuse gamma rays towards the Galactic center,
which can be interpreted  by the annihilation of dark matter (DM). 
We investigate the necessary conditions for 
a consistent explanation for the two photon excesses, 
especially the requirement on the width-to-mass ratio $\Gamma/M$ and
$\phi$ decay channels, 
in a collection  of DM models 
where the DM particle can  be  scalar, fermionic and vector, and
$\phi$ can be generated  through
$s$-channel $gg$ fusion  or 
$q\bar q$ annihilation.
We show that 
the minimally required $\Gamma/M$ is determined by
a single parameter  proportional to $(m_{\chi}/M)^{n}$, 
where the integer $n$ depends  on 
the nature of the DM particle.
We find that
for the scalar DM model with $\phi$ generated from $q\bar q$ annihilation, 
the minimally required  $\Gamma/M$ can be as low as 
$\mathcal{O}(10^{-3})$.
For the scalar DM model with $\phi$ generated from $gg$ fusion and 
fermionic DM model with $\phi$ from $q\bar q$ annihilation, 
the required  $\Gamma/M$ are typically of $\mathcal{O}(10^{-2})$.
The vector DM models, however,  
require very large $\Gamma/M$ of order one.
For the DM models which can consistently explain both the excesses, 
the predicted cross sections for gamma-ray line are typically of 
$\mathcal{O}(10^{-31}-10^{-29})~\text{cm}^{3}\text{s}^{-1}$, 
which are close to  the current limits from the  Fermi-LAT experiment.
\end{abstract}

\newpage
\section{Introduction}
Recently,  the ATLAS and CMS collaborations have reported 
the results of the LHC Run-2 at  
center-of-mass energy $\sqrt{s}=13$~TeV,
based on the  integrated luminosity of 
3.2~fb$^{-1}$ and 3.3~fb$^{-1}$, %
respectively
\cite{LHC-diphoton}.
Both the collaborations have shown 
a possible excess in the events containing two photons,
suggesting the existence of a new $s$-channel resonance particle $\phi$.
The distribution of the observed events at ATLAS  favours
a mass of the resonance $M\approx 750$~GeV, and
a width-to-mass ratio  $\Gamma/M\approx 0.06$ with
a local (global) significance of 3.9~$\sigma$ (2.3~$\sigma$).
In the assumption of  a narrow width, 
the corresponding local (global) significance is 3.6~$\sigma$ (2.0~$\sigma$).
The CMS collaboration has also reported 
a similar excess at $M \approx 760$~GeV  with 
a local (global) significance of $2.9\sigma$ ($<1 \sigma$),
and  the event distribution slightly favours  a narrow width. 
A combined analysis of the CMS Run-1 (8~TeV) and Run-2  data showed that
the local (global) significance of the diphoton excess  increases to 
3.4~$\sigma$ (1.6~$\sigma$) with 
the best-fit diphoton invariant mass close to  $750$~GeV
\cite{LHC-diphoton-CMS-1603}. 
If  the two photons arise directly from the decay of the resonance $\phi$,
the resonance  must be electrically neutral, and
its spin  can be  0 or 2 due to the Landau-Yang theorem
\cite{LangdauYang}.
Assuming  a large width, 
the ATLAS (CMS)  data favour a diphoton production cross section 
$10\pm 3~\mbox{fb}$ ($6\pm 3~\mbox{fb}$)
\cite{
Franceschini:2015kwy%
}. 
Other analyses assuming narrow width give
$\sim 6.2 \ (5.6)$~fb %
for ATLAS (CMS)
\cite{
Buttazzo:2015txu,%
Falkowski:2015swt%
}.

The LHC diphoton excess,  if  confirmed,
is a clear indication of  new physics beyond the standard model (SM).
Furthermore, $\phi$ is unlikely to  be the only new particle.
Since $\phi$ is electrically neutral, 
it can only couple to photons through loop processes. 
If the loops involve only the SM charged particles,
$\phi$ should decay into these SM particles with large rates,
as $\phi$ is much heavier than all the SM particles.
The corresponding production cross sections can easily  reach $\mathcal{O}(\mbox{pb})$  
which are  too large to escape the detection at LHC Run-1
(see e.g. 
\cite{
Low:2015qep,
Agrawal:2015dbf%
}).
If the large width $\Gamma/M\approx 0.06$ favoured  by ATLAS is confirmed,
the resonance $\phi$ is likely to have additional  tree-level invisible decays.
An intriguing  possibility is  that 
$\phi$ also couples to the dark matter (DM) particles 
which contribute to $\sim 26.8\%$ of the energy budget of our Universe.
In this scenario, 
$\phi$ plays the role of a portal connecting 
the invisible and visible world.
The excess of diphoton events suggests that 
the DM particle should at least couple to photons,
and  also couple to gluons or quarks depending on 
the production mechanism  of $\phi$ at the LHC.
The phenomenological implications  such as the DM relic density, 
DM direct and indirect detections have been extensive studied  
\cite{
Mambrini:2015wyu,
Backovic:2015fnp,
Franceschini:2015kwy,%
Bi:2015uqd,
Barducci:2015gtd,
Bai:2015nbs,
Han:2015cty,
Bauer:2015boy,
Dev:2015isx,
Davoudiasl:2015cuo,%
D'Eramo:2016mgv,%
Chu:2012qy,%
Park:2015ysf%
}.

If the DM particles can couple to the SM particles indirectly, 
the annihilation of the DM particles in the Galactic halo can generate 
extra  flux of cosmic-ray particles and photons.
Compared with the cosmic ray charged particles, 
the photons  are not deflected by the Galactic magnetic fields and 
do not loss energy during the propagation in the Galactic halo.
Thus they are of crucial importance in searching  for 
the signals  of halo DM annihilation.
The Galactic Center (GC) is expected to harbour high densities of DM,
as suggested by N-body simulations, 
which makes it a promising place to 
look for  photon signals of DM annihilation or decay.
Recently,
a number of groups including Ferm-LAT collaboration have  independently
found statistically strong evidence for  an excess in cosmic gamma-ray fluxes at 
energy $\sim2$~GeV towards
the inner regions  around the Galactic center (GC) from the data of Fermi-LAT
\cite{Goodenough:2009gk,
Hooper:2010mq,
Boyarsky:2010dr,
Abazajian:2010zy,
Hooper:2011ti,
Abazajian:2012pn,
Gordon:2013vta,
Macias:2013vya,
Abazajian:2014fta,%
Hooper:2013rwa,
Huang:2013pda,
Daylan:2014rsa,
Zhou:2014lva,%
Calore:2014xka,%
TheFermi-LAT:2015kwa%
}.
The morphology  of this GC excess (GCE) emission is consistent with 
a spherical emission profile expected from DM annihilation.
The origin of the GCE is still under debate.
There  exists  plausible astrophysical explanations 
such as the unresolved point sources of mili-second pulsars
\cite{
Abazajian:2010zy,%
Hooper:2011ti,%
Abazajian:2012pn,%
Gordon:2013vta,%
Bartels:2015aea,%
Lee:2015fea%
}
and
the interactions between the cosmic rays and the molecular gas
\cite{Macias:2013vya,%
Abazajian:2014fta,%
Gaggero:2015nsa%
}.
Halo DM annihilation can also provide a reasonable explanation.
The determined energy spectrum of the excess emission
although depending on the choices of  diffuse gamma-ray background templates, 
is in general compatible with  the scenario of  
$\sim40$~GeV DM particles self-annihilating into $b \bar b $ final states with 
a  cross section
$\langle \sigma v\rangle \approx (1-2)\times 10^{-26}\text{ cm}^{3}\text{s}^{-1}$
close to the typical thermal cross section for the observed DM relic aboundence
\cite{
Hooper:2013rwa,
Daylan:2014rsa%
}
(other possible final states were considered in Refs.
\cite{
Calore:2014xka,
Agrawal:2014oha,
Chu:2012qy%
}).

The possible connection between the LHC diphoton excess and 
the GCE was   first explored in
\cite{
Huang:2015svl%
}.
Assuming a pesudoscalar $\phi$ which couples dominantly  to 
$gg$, $\gamma\gamma$ and  scalar DM particles, 
it was shown that  the two reported photon excesses can be 
simultaneously explained
if the total width of $\phi$ is large  enough $\Gamma/M\gtrsim\mathcal{O}(10^{-2})$ which is  favoured by the current ATLAS data.
The phenomenological consequences of such a connection was further discussed in 
Refs.~\cite{
Hektor:2016uth%
} and 
\cite{
Krauss:2016cdi%
}.

A large total width of $\phi$, if confirmed, implies that 
the new physics sector is  strongly coupled, or 
the resonance $\phi$ has large number of  decay channels.
In this work, we investigate the  generic conditions for 
a consistent explanation for  possible the LHC diphoton excess and GCE,
especially the requirement on total width of $\phi$ in 
a wide range of DM models where
the DM particle can  be  scalar, fermionionic and vector, and
$\phi$ can be generated  by 
$s$-channel gluon fusion  or quark-antiquark annihilation at parton level.
We show that the minimally required $\phi$ width is determined by
a single parameter  proportional to $(m_{\chi}/M)^{n}$, 
where the integer $n$ depends  on 
the spins of the DM particle and its decay final states.
We find that
for scalar DM model with  $\phi$ generated from $q\bar q$ annihilation, 
the minimally required  $\Gamma/M$ can be as low as $\mathcal{O}(10^{-3})$.
For scalar DM model with $\phi$ generated from $gg$ fusion and 
fermionic DM model with $\phi$ from $q\bar q$ annihilation, 
the required  $\Gamma/M$ reaches $\mathcal{O}(10^{-2})$.
Other models such as the vector DM model requires larger $\Gamma/M$ of order one
which is already disfavoured by the current data.
For the same DM model, the required width of $\phi$ is always smaller
in $q\bar q$ channel than that in the $gg$ channel.
For the DM models which can simultaneously account for the diphoton excess and the GCE, 
the predicted cross sections for gamma-ray line are typically of 
$\mathcal{O}(10^{-30})~\text{cm}^{3}\text{s}^{-1}$, 
which is close to  the current limits imposed by the  Fermi-LAT data.
These models can be distinguish by LHC and Fermi-LAT in the near future.

The rest of this paper is organized as follows. 
In section 2,
we overview the interpretation of the diphoton excess, and 
derive model-independent conditions for a consistent explanation for 
the diphoton excess and the GCE.
In section 3,
we discuss model-independently the implications of the GCE for the DM properties.
In section 4, we determine the allowed parameters in various  DM models in which
the DM particles can be scalar, fermionic and vector with $\phi$ generated by $gg$ fusion
and $q\bar q$ annihilation.
The conclusion is given in section 5.

\section{The LHC diphton excess}
We consider the simplest scenario where
the diphoton events are produced from
the decay of  the $s$-wave resonance $\phi$  which 
is generated through $X\bar{X}$ fusion or annihilation process, 
where  $X\bar X=gg$, $\gamma\gamma$ and $q\bar q$ ($q=u,d,c,s,t,b$).
The production cross section for 
the process $pp\to\phi\to\gamma\gamma$ in 
the narrow-width approximation is given by 
\begin{align}\label{eq:diphoton}
\sigma_{\gamma\gamma}=\frac{2J+1}{(\Gamma/M)s}
\left(\sum_{X}C_{X\bar{X}}\frac{\Gamma_{X\bar{X}}}{M}\right)
\left(\frac{\Gamma_{\gamma\gamma}}{M}\right),
\end{align}
where 
$J$ is the spin of $\phi$, and 
the coefficients $C_{X\bar{X}}$ incorporate the integration  over 
the parton distribution functions  of the protons. 
For instance,
at the center-of-mass energy $\sqrt{s}=13\ (8)$ TeV, 
$C_{gg}\approx2137\ (174)$,
$C_{b\bar{b}}\approx15.3\ (1.07)$ and 
$C_{c\bar{c}}\approx36\ (2.7)$
\cite{Franceschini:2015kwy}.
Higher order QCD corrections can be taken into account by including
the $K$-factors with typical values $K_{gg\ (q\bar{q})}\approx1.48\ (1.20)$.
For the sake of simplicity, 
we consider the case where $\phi$ is spin zero, and
one channel of $X\bar X$ dominates the $\phi$ production at a time.
The process of $\gamma\gamma$ fusion is always included, 
as it is irreducible. 
In the limit of $\Gamma_{X\bar X}\gg \Gamma_{\gamma\gamma}$,
the values of the partial decay widths required to account for 
the diphoton excess at Run-2  are estimated as
\begin{align}
\left(\frac{\Gamma_{X\bar{X}}}{M}\right)
\left(\frac{\Gamma_{\gamma\gamma}}{M}\right)
\approx \frac{2.1\times 10^{-4}}{C_{X\bar{X}}}
\left(\frac{\sigma_{\gamma\gamma}}{8~\mbox{fb}}\right)
 \left(\frac{\Gamma/M}{0.06}\right) .
\end{align}
The non-observation of any excess at Run-1 (8~TeV) 
already imposes stringent limits on the cross sections for
a number of final states generated from the decay of a generic resonance
\begin{equation}\label{eq:run-1-limits}
\begin{tabular}{ccc}
$\sigma_{Z\gamma} \leq 4.0~\text{fb}$\cite{Aad:2014fha},   
$\sigma_{ZZ} \leq 12~\text{fb}$\cite{Aad:2015kna},  
$\sigma_{WW} \leq 40~\text{fb}$\cite{Khachatryan:2015cwa,Aad:2015agg}, 
\\
$\sigma_{\gamma\gamma} \leq 1.5~\text{fb}$
\cite{
Aad:2015mna%
},
$\sigma_{jj} \leq 2.5~\text{pb}$\cite{Aad:2014aqa},  
$\sigma_{b\bar b} \leq 1.0~\text{pb}$
\cite{
Khachatryan:2015tra%
},  
\end{tabular}
\end{equation}
The enhancement of the production cross section at Run-2 relative to 
that at Run-1 can be  described by the gain factor
$r=\sigma_{\text{13TeV}}/\sigma_{\text{8TeV}}
\approx 0.38 C_{X\bar X}(13\text{TeV})/C_{X\bar X}(8\text{TeV})$.
In order to account for an excess seen at Run-2 but not Run-1,
a large value of $r$ is favoured.
The production channels with leading $r$ factors are 
$r_{b\bar{b}}\approx 5.4$, $r_{gg}\approx 4.7$,  and 
$r_{c\bar{c}}\approx 5.1$.
Other channels have smaller gain factors, 
for instance,
$r_{ss}\approx 4.3$, 
$r_{dd} \approx 2.7$, 
$r_{uu}\approx 2.5$,  and 
$r_{\gamma\gamma} \approx 1.9$.
Thus they are not considered further in this work.
In the case where $\phi$ also couples to DM particles,
the total width of $\phi$ is given by 
\begin{align}\label{eq:total-width}
\Gamma & =\Gamma_{gg(q\bar{q})}+\kappa\Gamma_{\gamma\gamma}+\Gamma_{\chi\chi},
\end{align}
where the factor $\kappa=(1+\Gamma_{ZZ}/\Gamma_{\gamma\gamma}+\Gamma_{Z\gamma}/\Gamma_{\gamma\gamma}+\Gamma_{WW}/\Gamma_{\gamma\gamma})$
absorbs  the contributions from 
$ZZ$, $Z\gamma$ and $WW$ final states, 
which depends on the couplings 
between $\phi$  and  the SM weak gauge bosons in a given model.
If the total width $\Gamma$ can be determined by the experiment,
\eq{eq:total-width} can place  an important constraint on 
the properties of the DM particle.

If the diphoton events are generated dominantly by 
the process of gluon fusion (quark-antiquark annihilation)
$gg\ (q\bar q) \to \phi \to \gamma\gamma$,
the cross sections of diphoton  production  and DM annihilation are
strongly correlated,
as the DM particles inevitably annihilate into these states through the 
same intermediate state,
$\chi\bar\chi \to \phi\to gg, (q\bar q), \gamma \gamma$.

The same DM annihilation process determines 
both the DM relic density and the DM indirect detection signals. 
For the $s$-channel DM annihilation process $\chi\chi\to\phi\to X\bar{X}$, 
the corresponding thermally-averaged 
product of the  DM annihilation cross section and the DM relative velocity can be 
written in a generic  form
\begin{align}\label{eq:sigmav}
\langle\sigma v\rangle_{X\bar{X}} & =\frac{8\pi\eta_{\chi}R_{X\bar{X}}}{s_{\chi}^{2}\left[\left(1-4m_{\chi}^{2}/M^{2}\right)^{2}+(\Gamma/M)^{2}\right]m_{\chi}^{2}\beta_{\chi}(M^{2})}\left(\frac{\Gamma_{\chi\chi}}{M}\right)\left(\frac{\Gamma_{X\bar X}}{M}\right),
\end{align}
where 
$\eta_{\chi}=2\ (1)$ for the DM particle (not) being its own antiparticle,
$s_{\chi}$ is the spin degrees-of-freedom of the DM particle with 
$s_{\chi}=$1, 2 and 3 for the DM being a  scalar,  fermion and vector, 
respectively. 
The quantity $\beta_{X}(s)\equiv(1-4m_{X}^{2}/s)^{1/2}$
is the velocity of the particle $X$ from the decay $\phi^{(*)}\to X\bar{X}$
with a squared center-of-mass energy $s$. 
The function $R_{X\bar{X}}$  is essentially 
the ratio of $\phi$ decay squared amplitudes at $s\approx4m_{\chi}^{2}$ and $M^{2}$ 
\begin{align}
R_{X\bar{X}}(m_{\chi}^{2}/M^{2}) & =\frac{\sum|M_{\phi\to\chi\chi}(s=4m_{\chi}^{2})|^{2}\sum|M_{\phi\to X\bar{X}}(s=4m_{\chi}^{2})|^{2}\beta_{X}(4m_{\chi}^{2})}{\sum|M_{\phi\to\chi\chi}(s=M^{2})|^{2}\sum|M_{\phi\to X\bar{X}}(s=M^{2})|^{2}\beta_{X}(M^{2})}.
\end{align}
For a consistent explanation of the diphoton excess and the GCE,
\eqs{eq:diphoton}, (\ref{eq:total-width}) and (\ref{eq:sigmav}) must be satisfied simultaneously.
The corresponding solutions for the $\phi$ partial decay widths 
in the limits $m_{\chi}/M\ll 1$ and $\Gamma/M\ll 1$
are given by
\begin{align}\label{eq:twoSolutions}
\left(\frac{\Gamma_{X\bar X}}{M}\right)=\frac{1}{2}\left[\left(\frac{\Gamma}{M}\right)\pm\Delta^{1/2}\right], & \left(\frac{\Gamma_{\text{\ensuremath{\gamma\gamma}}}}{M}\right)=\frac{\sigma_{\gamma\gamma}s(\Gamma/M)}{C_{X\bar X}(\Gamma_{X\bar X}/M)}  ,
\end{align}
where
\begin{align}\label{eq:delta}
\Delta  
\equiv
\left(\frac{\Gamma}{M}\right)^{2}
-4\left[
\frac{s_{\chi}^{2}m_{\chi}^{2}\beta_{\chi}(M^{2})\langle\sigma v\rangle_{X\bar X}}{8\pi\eta_{\chi}R_{X\bar X}}
+\frac{\kappa\sigma_{\gamma\gamma}s}{C_{X\bar X}}\frac{\Gamma}{M}
\right]  .
\end{align}
The necessary condition for the existence of the solutions is $\Delta \geq 0$. 
As it can be seen in the following sections,
in most DM models $R_{X\bar X}\propto (m_{\chi}/M)^{2n},\ (n=1,2,3,\dots)$.
Since we are interested in the case of GCE where 
$m_{\chi}\ll M$,
the second term in the square brackets in \eq{eq:delta} can be
safely neglected. 
In a good approximation, the condition can be written as
\begin{align}\label{eq:totalWidthLimit}
\frac{\Gamma}{M}  
\gtrsim
\left(
\frac{s_{\chi}^{2}m_{\chi}^{2}\beta_{\chi}\langle\sigma v\rangle_{X\bar X}}{2\pi\eta_{\chi}R_{X\bar X}}  
\right)^{1/2}  .
\end{align}
For a given DM model,
the factor $R_{X\bar X}$ is fixed.
If the diphoton excess is consistent with the DM thermal relic density 
which is set by DM annihilation into $X\bar X$,
then the annihilation cross section must be close to 
the typical thermal cross section
$\langle\sigma v\rangle_{X\bar{X}}
\approx 
\langle\sigma v\rangle_{F}
=
3\times 10^{-26}\text{ cm}^{3}\text{s}^{-1}$.
From the value of  $\Gamma/M$ determined by the experiment, 
one can derived an upper limit on  $\langle\sigma v\rangle_{X\bar{X}}$ 
as a function of $m_{\chi}$ from \eq{eq:totalWidthLimit},
which depends on the nature of the DM particle and 
the final state $X\bar X$.
On the other hand,
if  the diphoton excess is required to be consistent with the GCE, 
since both $\langle\sigma v\rangle_{X\bar{X}}$ and 
$m_{\chi}$ can be determined by the GCE data,
\eq{eq:totalWidthLimit} can lead to a minimal requirement on 
the total width $\Gamma/M$.

The diphoton excess suggests that 
the DM particles inevitably annihilate into two-photon final states through 
$s$-channel $\phi$ exchange, 
which results in a spectral line in the generated gamma-ray flux with 
photon energy centered  at $E_{\gamma}=m_{\chi}$. 
The spectral line is difficult to be mimicked by conventional astrophysical
contributions, if observed, can be a strong evidence for halo DM annihilation or decay.
If the diphoton excess is generated from $X\bar X$ initial states, 
from \eq{eq:diphoton} and (\ref{eq:sigmav}), it follows that
\begin{align}
\langle\sigma v\rangle_{\gamma\gamma}
=\frac{\sigma_{\gamma\gamma}}{\sigma_{X\bar X}} 
\frac{R_{\gamma\gamma}}{R_{X\bar X}}
 \langle\sigma v\rangle_{X\bar X}  ,
\end{align}
where $\sigma_{X\bar X}$ is the cross section for 
the production of $X\bar X$ final states  through 
intermediate state $\phi$ from $X\bar X$ fusion or annihilation at the LHC, 
i.e. $X\bar X\to \phi \to X\bar X$.
For a given DM model, 
the values of $R_{\gamma\gamma}/R_{X\bar X}$ is fixed.
Thus, from the Run-1 upper limit on $\sigma_{X\bar X}$, 
one can obtain a $lower$ limit on $\langle\sigma v\rangle_{\gamma\gamma}$.
It was shown in Ref.~\cite{Huang:2015svl} that
a lower limit of $\langle \sigma v \rangle_{\gamma\gamma}\gtrsim ~4.8\times 10^{-30}~\mbox{cm}^{3}\mbox{s}^{-1} $ can be obtained in a scalar DM model with $\phi$ generated through $gg$
fusion.

If $\phi$ is allowed to  couple to $Z\gamma$,
the DM annihilation can generate a gamma-ray line with 
photon energy at
$E_{\gamma}  = m_{\chi} (1-m_{Z}^{2}/4m_{\chi}^{2})$.
The annihilation cross section for the process $\chi\chi\to\phi\to Z\gamma$ is related to 
that for $\chi\chi\to\phi\to \gamma\gamma$ as follows
\begin{align}
\langle\sigma v\rangle_{Z\gamma}
=
\frac{\sigma_{Z\gamma}}{\sigma_{\gamma\gamma}}
\frac{\tilde\beta^{6}_{Z}(4m^{2}_{\chi})}{\tilde\beta^{6}_{Z}(M^{2})}
\langle\sigma v\rangle_{\gamma\gamma}  ,
\end{align}
where $\tilde\beta_{X}(s)=(1-m_{X}^{2}/s)^{1/2}$.
Since $\sigma_{Z\gamma}/\sigma_{\gamma\gamma}$ is a known in a give model,
a lower limit on $\langle \sigma v \rangle_{Z\gamma}$ can be obtained
in a similar way.

\section{The Galactic Center Excess}
The annihilation of DM particles into $X\bar X$ final states generates 
diffuse gamma rays with a broad energy spectrum due to hadronization, 
while the annihilation into $\gamma\gamma$ generates 
a line-shape spectrum with energy centered at the DM particle mass. 
Both the signatures are under active searches by 
the current DM indirect detection experiments. 
The differential gamma-ray flux,
averaged over a solid angle $\Delta\Omega$ is given by 
\begin{align}\label{eq:fluxAvg}
\frac{d\Phi}{dE}=
\frac{\eta_{\chi} \rho_{0}^{2} r_{\odot} }{16\pi }
\frac{\langle\sigma v\rangle}{m_{\chi}^{2}}
\frac{dN_{\gamma}}{dE} J ,
\end{align}
where 
$r_{\odot} \approx 8.5$~kpc is the distance from the Sun to the GC, 
$\rho_{0} \approx 0.4\mbox{ GeV}/\mbox{cm}^{3}$ is 
the local DM density in the solar neighbourhood, and
$dN_{\gamma}/dE$ is the gamma-ray spectrum per DM annihilation.
The dimensionless $J$-factor which contains the information of 
DM density distribution is given by
\begin{align}
J=\int\frac{d\Omega}{\Delta\Omega}\int_{\text{l.o.s}}
\left( \frac{\rho(r)}{\rho_0} \right)^{2}
\frac{ds}{r_{\odot}} ,
\end{align}
where
$\rho(r)$ is the spatial distribution of halo DM energy density,
with  $r$ the distance to the GC. 
The integration is to be performed over the distance $s$
along the light-of-sight which is related to $r$ through the relation
$r^{2}=r_{\odot}^{2}+s^{2}-2sr_{\odot}\cos\psi$, where $\psi$ is
the angle of the direction away from the GC. 
N-body simulations suggest an universal DM density profile of 
the Navarro-Frenk-White (NFW)  form \cite{Navarro:1996gj} 
\begin{align}
\rho(r)=\rho_{s}\left(\frac{r}{r_{s}}\right)^{-\gamma}\left[1+\left(\frac{r}{r_{s}}\right)^{\alpha}\right]^{\frac{\gamma-\beta}{\alpha}}  ,
\end{align}
which is characterized by the parameters $\alpha$, $\beta$, $\gamma$,
and a reference scale $r_{s}\simeq20$~kpc. For the standard NFW
profile, $\alpha=\gamma=1$ and $\beta=3$. 
The normalization factor $\rho_{s}$ is determined by 
the local DM density 
$\rho(r_{\odot})=\rho_{0}$.

\begin{figure}[!htbp]
\begin{center}
\includegraphics[width=0.45\textwidth]{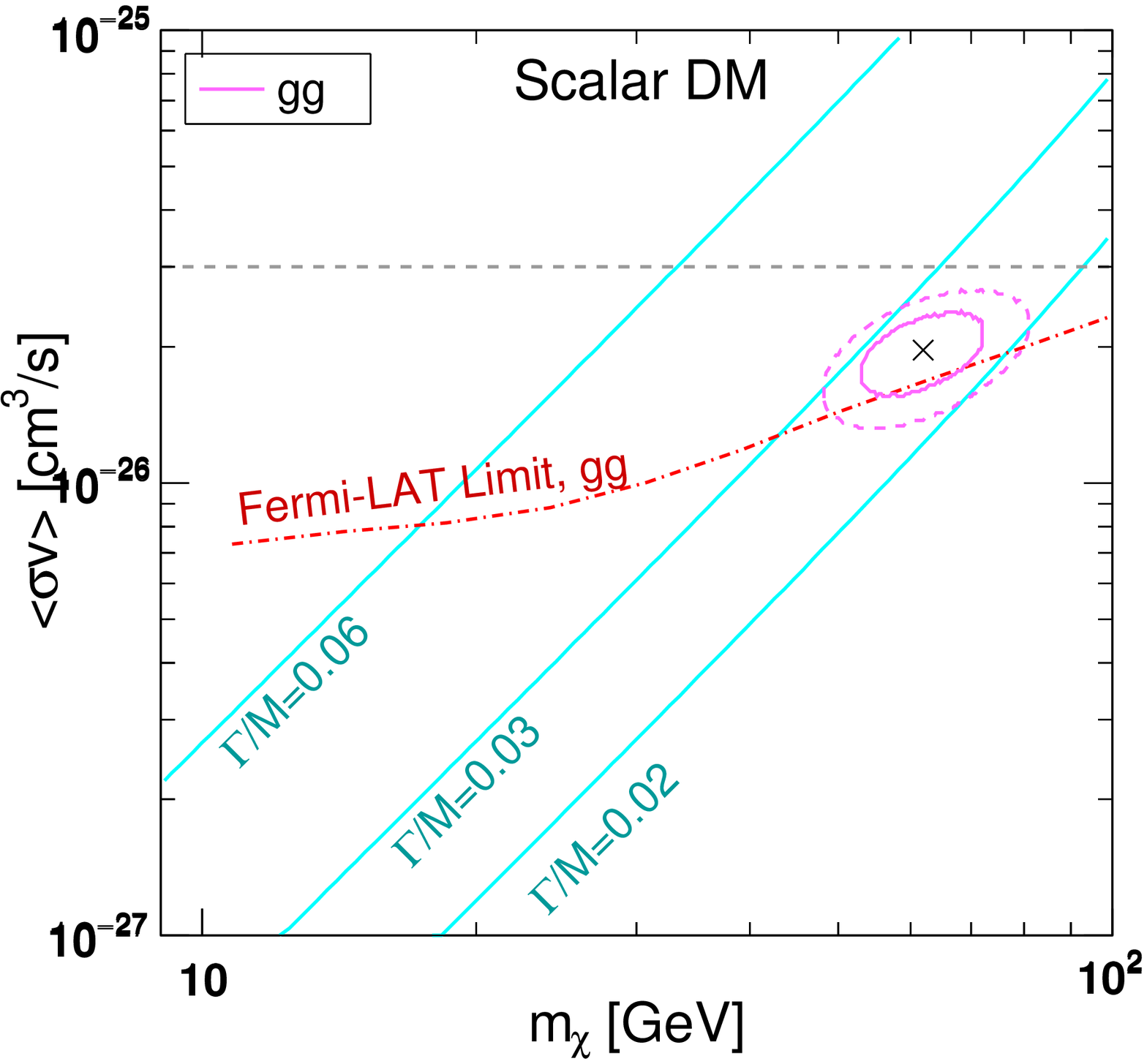}
\includegraphics[width=0.45\textwidth]{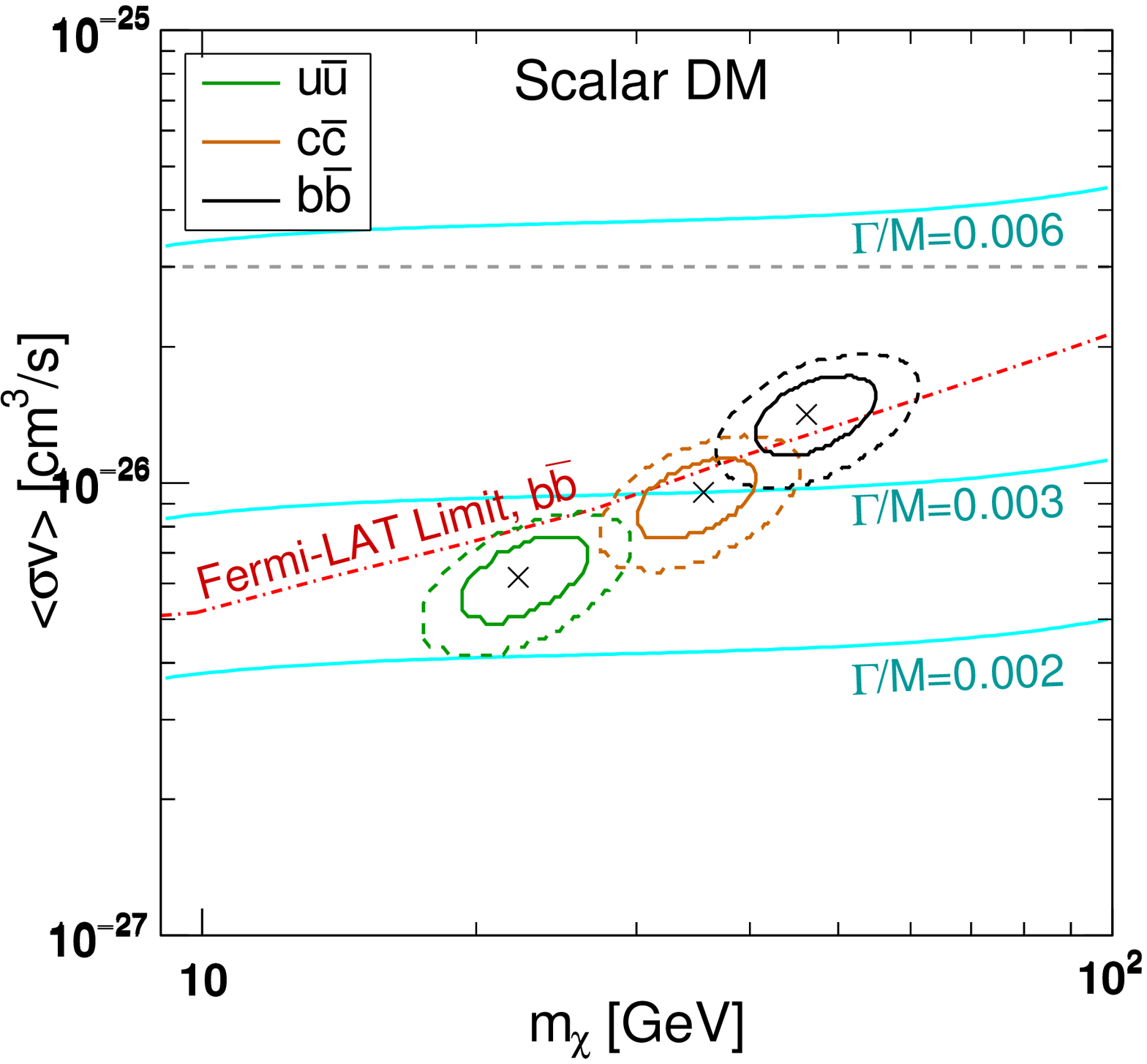}
\\
\includegraphics[width=0.45\textwidth]{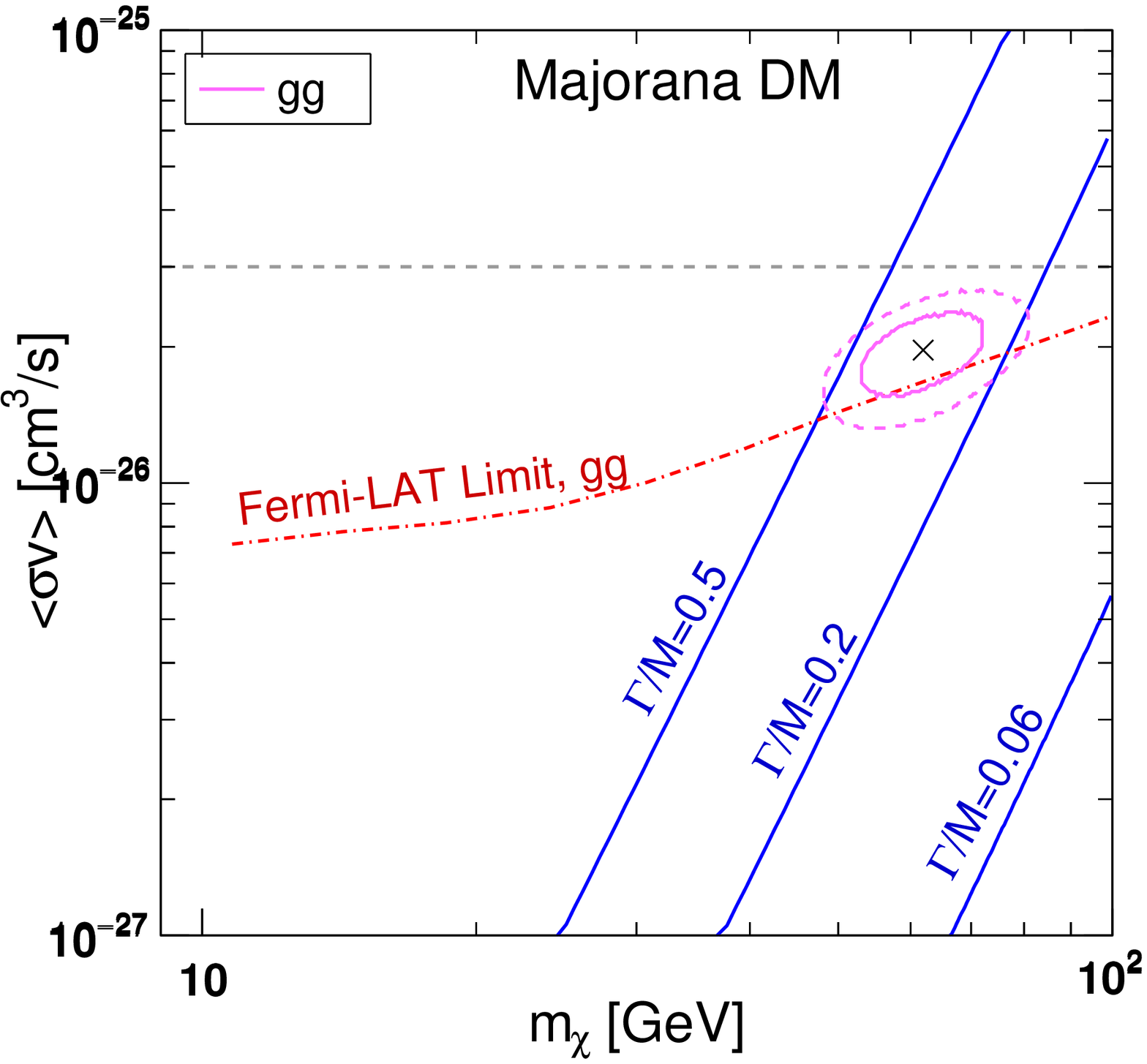}
\includegraphics[width=0.45\textwidth]{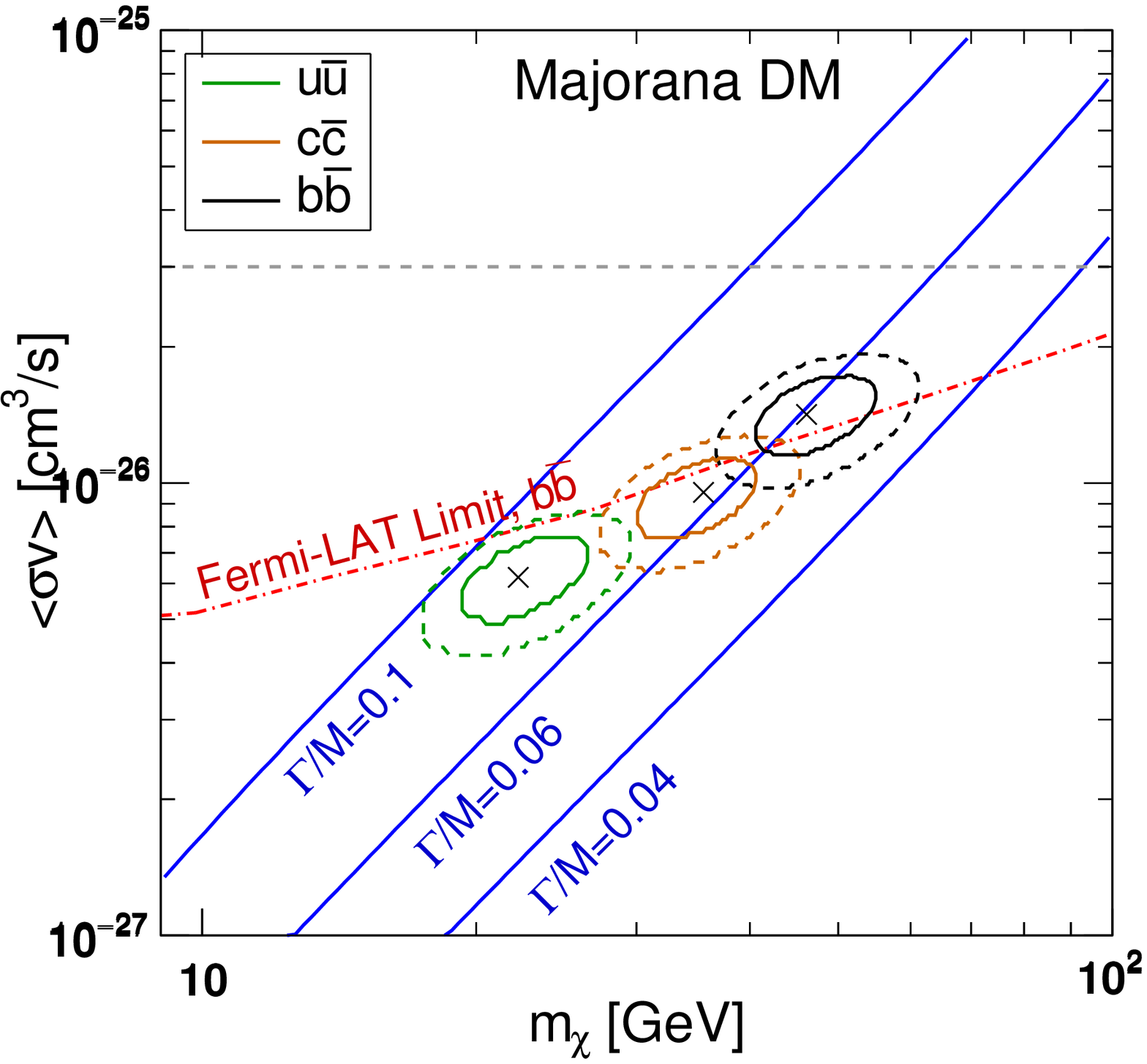}
\caption{
Upper panels)
Left: 
regions of DM mass and annihilation cross section allowed by the GCE
data at $68\%$ and $95\%$ C.Ls.  
for the scalar DM particle annihilating into 
$gg$ final states through $\phi$ exchange.
Upper limits on the  annihilation cross section as a function of DM particle mass
for fixed values of  $\Gamma/M$ are shown.
See text for detailed explanations.
The horizontal line indicates the typical thermal annihilation cross section of 
$3\times 10^{-26}~\text{cm}^{3}\text{s}^{-1}$.
The $95\%$ C.L. upper limits from the Fermi-LAT data on the gamma rays from dSphs 
\cite{
Ackermann:2015zua%
}
are also shown.  
The limits on $gg$ channel was derived using 
a conservative rescaling approach detailed in
\cite{
Huang:2015svl%
}. 
Right: the same as left but for DM annihilation into $b\bar b$, $c\bar c$ and $u\bar u$ final states.
Lower pannels)
The same as upper panels but for Majorana fermionic DM model.
}
\label{fig:GCEfit}
\end{center}
\end{figure}

We  determine the favoured values  of  
$m_{\chi}$ and $\langle\sigma v\rangle_{X\bar X}$ for 
a number of annihilation final states such as
$gg$, $b\bar b$, $c \bar c$ and $u\bar u$,
from fitting to  the GCE data derived in Ref.~\cite{Calore:2014xka}.
In total there are 24 data points.
The  spectra of the prompt gamma rays $dN_{\gamma}/dE$ for 
DM annihilating into  $X\bar X$ are generated
by the Monte-Carlo simulation package Pythia 8.201~\cite{Sjostrand:2007gs}. 
For the considered final states, 
the contributions from the inverse Compton scatterings can be safely neglected.
We choose a modified NFW profile with an inner slope $\gamma=1.26$, 
as suggested by the observed morphology of the gamma-ray emission 
\cite{Abazajian:2012pn,Gordon:2013vta,Hooper:2013rwa,Calore:2014xka}.
Making use of \eq{eq:fluxAvg}, 
the calculated diffuse gamma-ray fluxes are averaged over
a square region of interest (ROI) $20^{\circ}\times20^{\circ}$ in 
the sky with latitude $|b|<2^{\circ}$ masked out. 
The corresponding $J$-factor is $J=57.6$.
The best-fit DM particle masses and annihilation cross sections, 
and the corresponding $\chi^{2}$ and $p$-values are summarized  in \tab{tab:GCEdm}.
\begin{table}\begin{center}
    \begin{tabular}{ccccc}
    \hline\hline
    Channel  & $m_\chi$ (GeV)  & $\langle\sigma v\rangle_{bb}$  $(10^{-26}\text{cm}^3\text{s}^{-1}) $ & $\chi^2_{\text{min}}/\text{d.o.f.} $ & $p$-value \\
   \hline
    $b\bar{b}$    & $46.15^{+5.81}_{-3.53}$  & $1.42^{+0.18}_{-0.17}$ & 24.572/22  &  0.32 \\
    $c\bar{c}$    & $35.54^{+3.10}_{-4.12}$  & $0.95^{+0.12}_{-0.12}$ & 25.626/22 &  0.27 \\
    $u\bar{u}$    & $22.26^{+2.83}_{-1.91}$  & $0.62^{+0.10}_{-0.08}$ & 28.495/22 &  0.16 \\
    $gg$    & $ 62.01^{+6.56}_{-6.35}$  & $1.96^{+0.26}_{-0.24}$ & 24.665/22 &  0.31 \\
 \hline\hline
\end{tabular}
\caption{
Values of DM mass and annihilation cross sections determined from 
fitting to the GCE data. 
The DM particle is assumed to be its own antiparticle.
}\label{tab:GCEdm}
\end{center}\end{table}
In  \fig{fig:GCEfit}, we show the contours of the allowed regions for 
 the parameters $m_{\chi}$ and $\langle \sigma v\rangle_{X\bar X}$ at 
$68\%$ and $95\%$ C.L. for two parameters, 
corresponding to $\Delta \chi^{2}=2.3$ and 6.0, respectively.
As can be seen from the table, 
in the DM interpretation of the GCE, 
the required DM particle mass is in the range $\sim(20-70)$~GeV with
a cross section $(0.5-2)\times 10^{-26}\text{cm}^{3}\text{s}^{-1}$.
The most favoured channel is $b\bar b$. 
We emphasize that the $gg$ channel also gives reasonably good fit with 
a larger DM mass $\sim 60$~GeV, 
which is crucial for a consistent explanation with the diphoton excess,
as $gg$ fusion is also the favoured channel for 
the production of $\phi$ at the LHC Run-2.
These results are in good agreement with 
the previous analysis in Ref.~\cite{Calore:2014nla}.

At present, 
the most stringent constraints on the DM annihilation cross sections
are provide by the Fermi-LAT data on the diffuse gamma rays
of the dwarf spheroidal satellite galaxies (dSphs)
\cite{
Ackermann:2015zua%
}.
These limits are also shown in \fig{fig:GCEfit} for comparison purpose, 
where the limits on $gg$ channel was derived using 
a conservative rescaling approach detailed in Ref.
\cite{
Huang:2015svl%
}. 
It is known that there is an apparent tension between 
the GCE favoured regions and the Fermi-LAT limits.
Note that the DM velocity dispersion in the Galactic halo is 
quite different from that in the dSphs. 
The DM annihilation cross section favoured by the  GCE data and 
constrained  by the gamma rays of dSphs can only be compared under 
the  assumption that the cross section is velocity independent, 
which is in general not the case.
In the analysis of the Fermi-LAT collaboration, 
the uncertainties in the $J$-factors were taken into account assuming 
a  NFW type parametrization of the DM density profile. 
A recent analysis directly using the spherical Jeans equations rather than taking a
parametric DM density profile as input showed that
the $J$-factor can be smaller by a factor about $2-4$ for 
the case of Ursa Minor, 
which relaxes the constraints on the DM annihilation cross section to
the same amount
\cite{
Ullio:2016kvy%
}.

The annihilation of halo DM also generates  cosmic-rayparticles 
such as protons/antiprotons, electrons/positrons and neutrinos. 
Compared with  gamma-rays, 
the predictions for the flux of cosmic-ray charged particles  from DM annihilation suffer from 
large uncertainties in the  cosmic-ray propagation models.
For a DM particle mass below $\sim 100$~GeV, 
the predicted $\bar p/p$ ratio peaks at  lower energies below $\sim10$~GeV, 
which suffer from additional uncertainties due to the  solar activities.
The upper limits on the DM annihilation cross section 
from the AMS-02 and PAMELA data on $\bar p/p$ ratio for various channels 
have been studied  for variuos propagation models 
and DM density profiles ( see e.g. 
\cite{
Jin:2015sqa,%
Jin:2015mka,%
Hooper:2014ysa,
Jin:2012jn,%
Jin:2014ica,%
Jin:2013nta%
}). 
In general, the obtained limits are weaker than that derived from 
the gamma rays of dSphs.  
The constraints from the cosmic-ray positrons  depends strongly on 
the annihilation final states. 
For leptonic final states such as $e^{+}e^{-}$ and $\mu^{+}\mu^{-}$, 
the derived upper limits from the AMS-02 positron flux can reach the typical thermal cross section 
for DM particle mass below 50--100~GeV
\cite{ 
Ibarra:2013zia%
}.
But for hadronic final states such as $b\bar b$, the corresponding limits are rather weak,
typically at $\mathcal{O}(10^{-24})~\text{cm}^{3}\text{s}^{-1}$.
The $gg$ final state generates a  softer positron spectrum in comparison with 
the $b\bar b$ final states. 
Thus the corresponding limits are expected to be even weaker.

\section{DM Models} 

We focus on the scenario where 
the 750 GeV resonance $\phi$ is a pseudo-scalar particle. 
For a scalar resonance, the related couplings are severely constrained by 
the null results of the DM direct detection experiments.
The UV origins of the  pseudoscalar $\phi$ can be
axion-like particles from the breaking of the Peccei-Quinn symmetry
\cite{PQsym},
pesudo-Goldstone boson from composite Higgs models
\cite{
Gripaios:2009pe%
},
or from the extended Higgs sectors of the SM 
\cite{
Angelescu:2015uiz,
Han:2015qqj,
Huang:2015rkj,
Moretti:2015pbj,
Badziak:2015zez,
Han:2016bus,
Han:2016bvl,
Gopalakrishna:2016tku,
Ge:2016xcq%
}, 
or left-right symmetric models
\cite{
Dey:2015bur,
Dasgupta:2015pbr,
Borah:2016uoi,
Huong:2016kpa,
Dev:2015vjd,%
Guo:2011zze,Guo:2010sy,Guo:2010vy,Guo:2008si,Wu:2007kt
}.
If $\phi$ is a SM singlet and couples to the SM gauge bosons through 
vector-like heavy fermions which have 
small mixings with the SM fermions,
the constraints from the oblique parameters $S$ and $T$, 
the EW precision test,
and the flavor physics can be relaxed
\cite{
Ellis:2015oso%
}.
A pseudo-scalar does not mix with the SM Higgs boson, 
and is less constrained by the measured properties of the Higgs boson. 
For fermionic DM, its annihilation into SM particles through $s$-channel
pseudoscalar exchange is not suppressed by the low velocity dispersion of 
the halo DM. 
Since $\phi$ is a pseudo-scalar, 
the cross sections for DM-nucleus scattering through 
quarks or gluons within the nucleus are 
either velocity suppressed or vanishing,
which  easily relaxes  the stringent upper limits from 
various DM direct detection experiments.

We assume that $\phi$  can  couple to the SM quarks directly and 
the SM gauge fields indirectly through loop processes (see e.g. Refs.~\cite{
others%
}).
Since $\phi$ is much heavier than the electroweak (EW) scale, 
we start with EW gauge-invariant effective interactions up to dimension-five
\begin{align}
\mathcal{L}\supset &
\frac{1}{2}\partial_{\mu} \phi \partial^{\mu} \phi
-\frac{1}{2}M^{2}\phi^{2}
-i y_{q}\bar q \gamma^{5} q \phi
-\frac{g^{2}_{1}}{2\Lambda}\phi B_{\mu\nu}\tilde{B}^{\mu\nu}
\nonumber\\
& -\frac{g^{2}_{2}}{2\Lambda}\phi W_{\mu\nu}^{}\tilde{W}^{\mu\nu}
-\frac{g^{2}_{g}}{2\Lambda}\phi G_{\mu\nu}^{}\tilde{G}^{\mu\nu},
\end{align}
where for the gauge fields 
$\tilde{F}_{\mu\nu}=\frac{1}{2}\epsilon_{\mu\nu\alpha\beta}F^{\alpha\beta}$,
$g_{1,2,g}$ are the dimensionless effective coupling strengths, 
$y_{q}$ is the Yukawa coupling strength, and 
$\Lambda$ is a common energy scale. 
After the EW symmetry breaking,
the interaction terms involving physical EW gauge bosons 
$A$, $Z$ and $W$ are given by
\begin{align}
\mathcal{L}_{\text{gauge}}
\supset &
-\frac{g^{2}_{A}}{2\Lambda}\phi A_{\mu\nu}\tilde{A}^{\mu\nu}
-\frac{g^{2}_{Z}}{2\Lambda}\phi Z_{\mu\nu}\tilde{Z}^{\mu\nu}
-\frac{g^{2}_{AZ}}{2\Lambda}\phi A_{\mu\nu}\tilde{Z}^{\mu\nu}
\nonumber\\
&-\frac{g^{2}_{W}}{2\Lambda}\phi W_{\mu\nu}\tilde{W}^{\mu\nu}
-\frac{g^{2}_{g}}{2\Lambda}\phi G_{\mu\nu}\tilde{G}^{\mu\nu},
\end{align}
where the physical gauge  couplings $g_{A}$, $g_{Z}$, $g_{ZA}$ and 
$g_{W}$ are related to  that in the gauge basis as 
\begin{align}
g^{2}_{A} &=g^{2}_{1}c_{W}^{2}+g^{2}_{2}s_{W}^{2}, 
\quad
g^{2}_{Z}=g^{2}_{1}s_{W}^{2}+g^{2}_{2}c_{W}^{2}, 
\nonumber\\
g^{2}_{ZA} &=2s_{W}c_{W}(g^{2}_{2}-g^{2}_{1}), 
\quad
g^{2}_{W}=g^{2}_{2}
\end{align}
with $s_{W}^{2}=1-c_{W}^2=\sin^{2}\theta_{W}\approx 0.23$.
For the three extreme cases: $g_{1}=0$,
$g_{1}=g_{2}$ and $g_{2}=0$, the partial widths of $ZZ$, $Z\gamma$
and $WW$ relative to that of $\gamma\gamma$ and the values of $\kappa$
are listed in \tab{tab:couplings}.
We should focus on the case of $g_{2}=0$,
namely $\phi$ is not charged under the $SU(2)_{L}$ gauge group.
Note that the case of $g_{1}=0$ is severely constrained by the Run-I data on 
the $Z\gamma$ and $ZZ$ production rates, 
as it can be seen from \eq{eq:run-1-limits} and \tab{tab:couplings}.
The related phenomenology in the case of $g_{1}=g_{2}$ is similar to 
that in the case of $g_{2}=0$,
except that the DM annihilation into $Z\gamma$ is forbidden.
Thus there is no gamma-ray line generated from $Z\gamma$ final states.

The partial decay widths for $\phi$ decaying into the SM gauge bosons and 
the fermions are given by
\begin{align}\label{eq:widths}
\frac{\Gamma_{\gamma\gamma}}{M}
&=
\pi\alpha_{A}^{2}\left(\frac{M}{\Lambda}\right)^{2}, \
\frac{\Gamma_{gg}}{M}=8\pi\alpha^{2}_{g}\left(\frac{M}{\Lambda}\right)^{2},
\ 
\frac{\Gamma_{q\bar q}}{M}=\frac{3y_{q}^{2}\beta_{q}(M^{2})}{8\pi} ,
\end{align}
respectively, 
where $\alpha_{A,g}=g^{2}_{A,g}/4\pi$. 
For the spin nature of DM particles, 
we consider three classes of models
where the DM particles can be scalar, fermionic and vector.

\begin{table}\begin{center}
\begin{tabular}{ccccc}
\hline\hline
models & $\Gamma_{ZZ}/\Gamma_{\gamma\gamma}$ & $\Gamma_{Z\gamma}/\Gamma_{\gamma\gamma}$ & $\Gamma_{WW}/\Gamma_{\gamma\gamma}$ & $\kappa$\tabularnewline
\hline 
$g_{1}=0$ & 10 & 6.4 & 35 & 53\tabularnewline
\hline 
$g_{1}=g_{2}$ & 0.9 & 0 & 1.9 & 3.8\tabularnewline
\hline 
$g_{2}=0$ & 0.081 & 0.57 & 0 & 1.7 \tabularnewline
\hline\hline 
\end{tabular}
\caption{
Ratios of $\phi$ decay widths 
$\Gamma_{ZZ}/\Gamma_{\gamma\gamma}$,
$\Gamma_{Z\gamma}/\Gamma_{\gamma\gamma}$,
$\Gamma_{WW}/\Gamma_{\gamma\gamma}$
and the value of $\kappa$ defined in \eq{eq:total-width} for 
three cases of $\phi$ couplings with the SM gauge bosons, 
$g_{1}=0$, $g_{1}=g_{2}$ and $g_{2}=0$, respectively.
}\label{tab:couplings}
\end{center}\end{table}

\subsection{Real scalar DM}
In the real scalar DM model, 
the Lagrangian for the DM particle $\chi$ and its interaction with  $\phi$ is given by
\begin{align}
\mathcal{L}  
\supset
\frac{1}{2} \partial_{\mu} \chi \partial^{\mu} \chi
-\frac{1}{2}m_{\chi}^{2}\chi^{2} 
-\frac{1}{2}g_{\chi}\phi\chi^{2}  ,
\end{align}
where $g_{\chi}$ is a dimensionful coupling strength, and
we have only included the most relevant interaction term.
Other possible interaction terms such as $\lambda\phi^{2}\chi^{2}/4$
are neglected  by assuming small couplings.

For  DM annihilation into $gg$ final states, 
the corresponding factor $R_{gg}$ defined in \eq{eq:sigmav} is given by 
\begin{align}
R_{gg}=16\left(\frac{m_{\chi}}{M}\right)^{4}  ,
\end{align}
which is typically of $\mathcal{O}(10)^{-4}$ for $m_{\chi}\approx 60$~GeV.
In this case, 
\eq{eq:totalWidthLimit} can be rewritten as
\begin{align}
\left(\frac{\Gamma}{M} \right)_{\text{scalar},gg}  
\geq
\frac{\beta_{\chi}^{1/2}(M^{2})\langle \sigma v\rangle_{gg}^{1/2} M^{2}}{8\pi^{1/2}m_{\chi}}  .
\end{align}
For a given value of $\Gamma/M$, 
the above inequality can be interpreted as 
the upper limit on $\langle \sigma v\rangle_{gg}$ as a function of $m_{\chi}$.
In the upper-left panel of \fig{fig:GCEfit}, 
we show this relation for three choices of 
$\Gamma/M=0.06$, 0.03 and 0.02, respectively.
If $\langle \sigma v\rangle_{gg}$ is required  to meet the thermal value $\langle \sigma v\rangle_{F}$, 
it can be seen that the DM particle mass has to be 
larger than $\sim 65$ GeV (90 GeV) for  $\Gamma/M=0.03$ (0.02).
While for  $\Gamma/M=0.06$, the constraint on the DM particle mass is rather weak. 

If $\Gamma/M$ is not fixed,
using the best-fit values of $m_{\chi}$ and $\langle \sigma v\rangle_{gg}$ for 
$gg$ channel  from \tab{tab:GCEdm},
a lower limit on the required total width of $\phi$ can be obtained as follows
\begin{align}
\left(\frac{\Gamma}{M} \right)_{\text{scalar},gg}
\gtrsim 
0.026 
\left( \frac{M}{750~\text{GeV}}\right)^{2}
\left( \frac{62~\text{GeV}}{m_{\chi}}\right)
\left( \frac{\langle \sigma v\rangle_{gg}}{2.0\times 10^{-26}~\mbox{cm}^{3}\mbox{s}^{-1}}\right)^{1/2}  .
\end{align} 
Thus the GCE  required typical minimal width-to-mass ratio is 
quite large of $\mathcal{O}(10^{-2})$, 
which is currently favoured by ATLAS, and
can be confirmed or ruled out soon by the upcoming LHC updated results.
Assuming $\phi$ is generated dominantly by $gg$ fusion,
a combined fit to both the data of diphton excess and GCE in 
this model has been carried out in Ref.
\cite{
Huang:2015svl%
}, 
which showed that
the total width of $\phi$ is dominated by $\Gamma_{\chi\chi}$, and 
the favoured partial widths $\Gamma_{gg}/M$ and $\Gamma_{\gamma\gamma}/M$ are of $\mathcal{O}(10^{-3})$ and $\mathcal{O}(10^{-5})$, respectively.

\begin{figure}[htb]
\begin{center}
\includegraphics[width=0.45\textwidth]{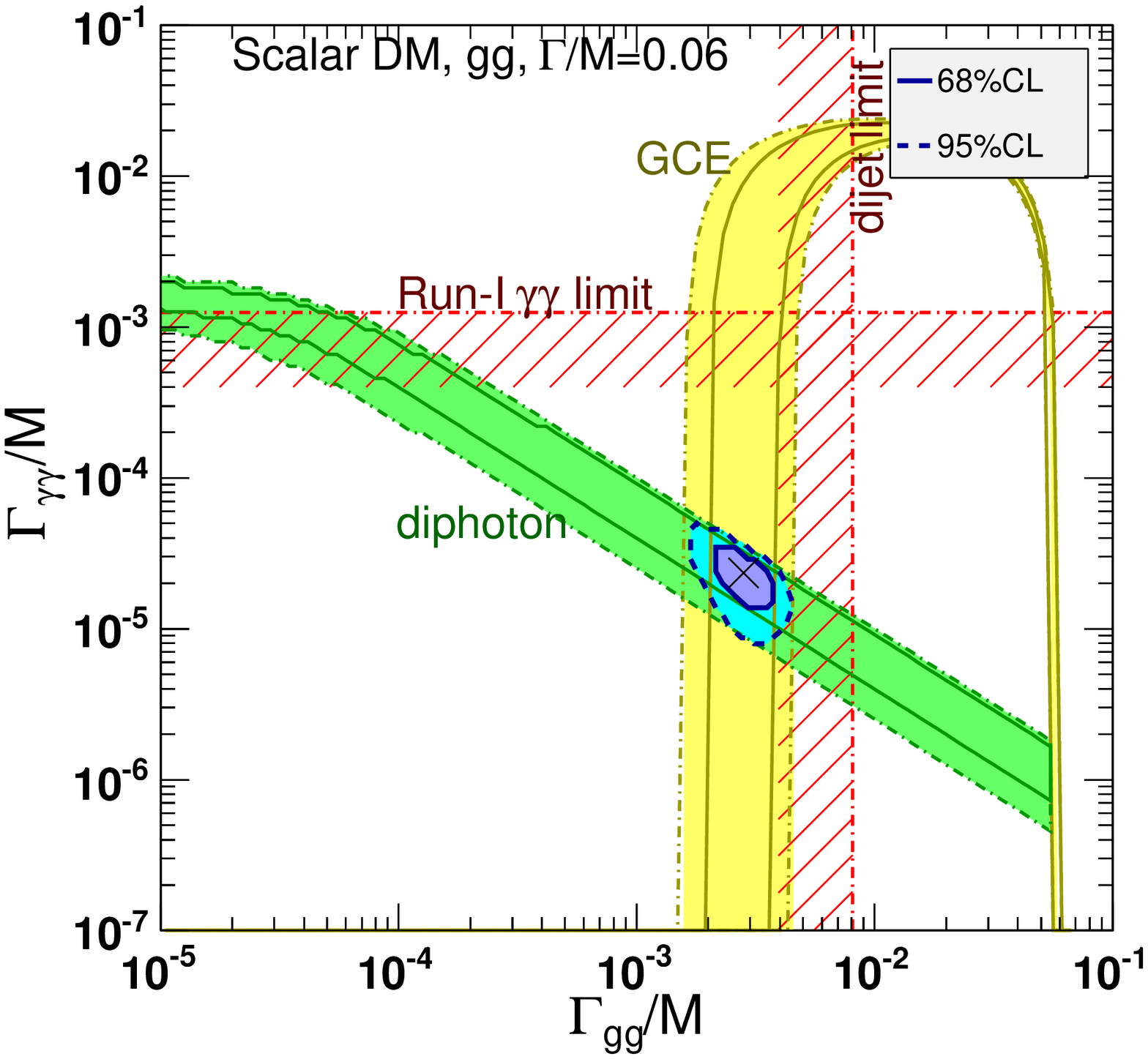}
\includegraphics[width=0.45\textwidth]{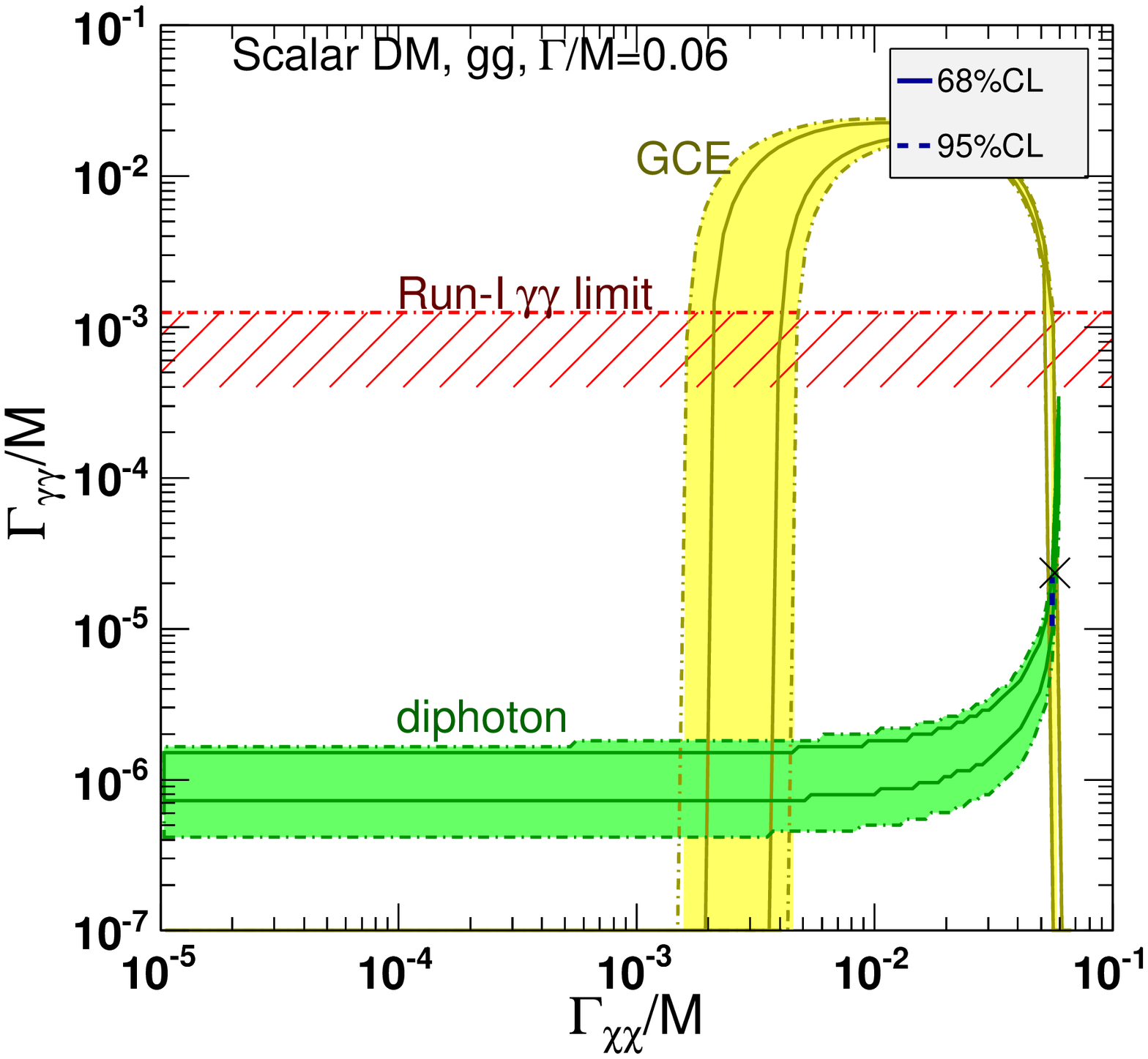}
\\
\includegraphics[width=0.45\textwidth]{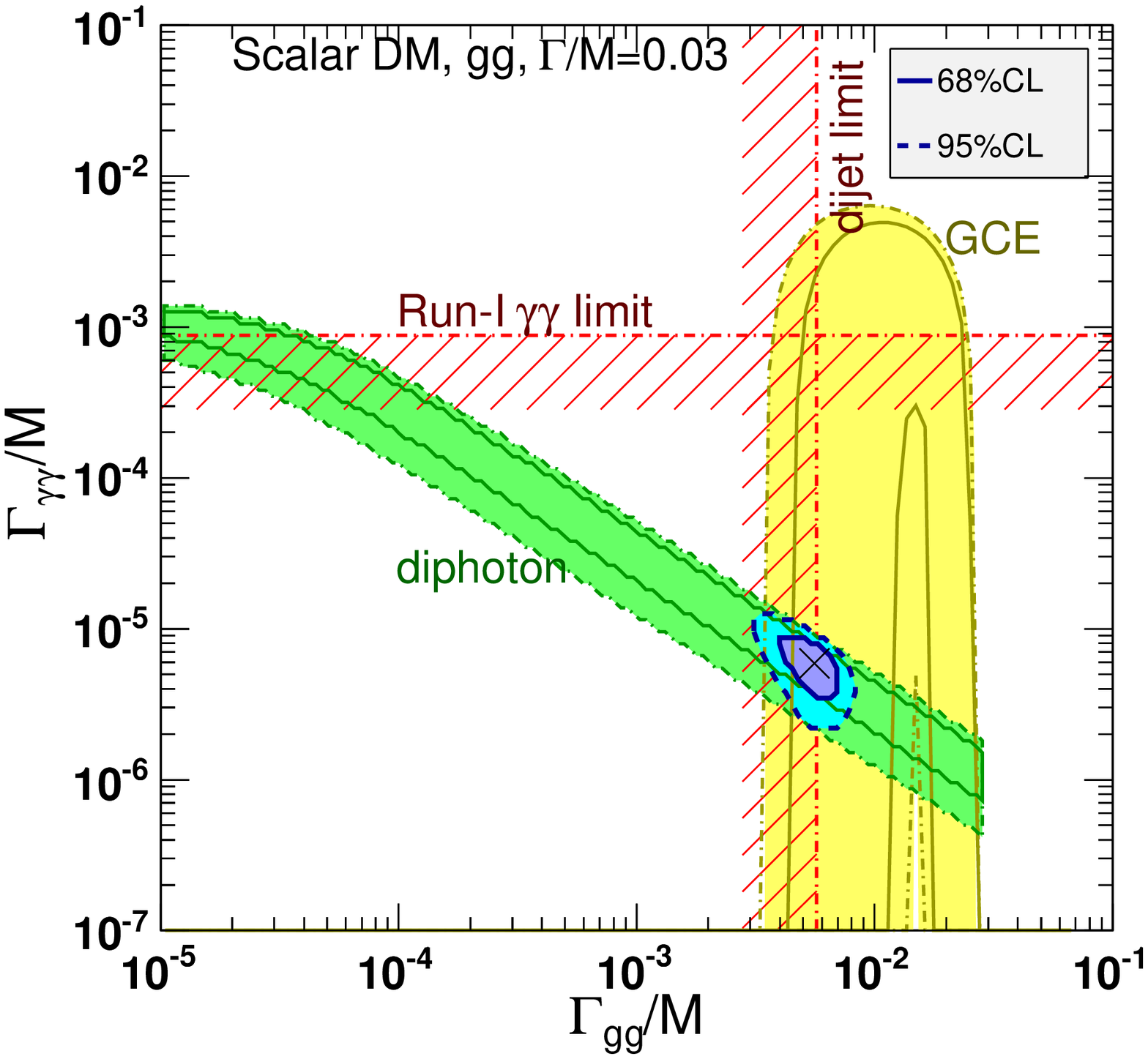}
\includegraphics[width=0.45\textwidth]{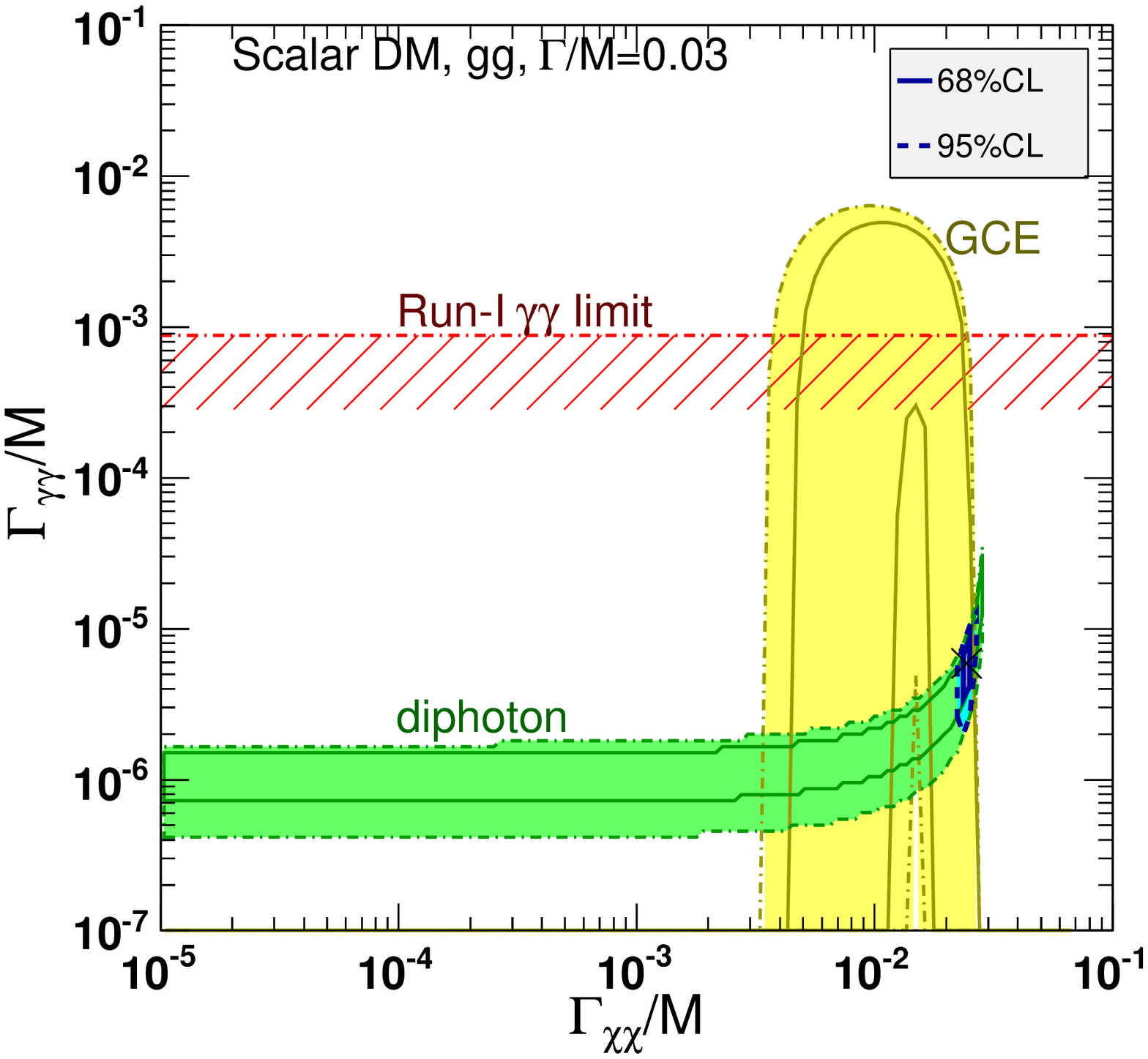}
\caption{
Upper panels)
Left:  Allowed regions at
$60\%$ and $95\%$ C.L. in 
$(\Gamma_{gg}/M, \Gamma_{\gamma\gamma}/M)$ plane from 
a combined fit to both the LHC diphoton excess and the GCE in 
a scalar DM model with $gg$ fusion,
together with the regions allowed by each individual experiment.
The upper limits from Run-1 on the dijet and diphoton production
cross sections are also shown.
The total width is fixed at $\Gamma/M=0.06$.
Right: The same as upper-left, 
but in $(\Gamma_{\chi\chi}/M, \Gamma_{\gamma\gamma}/M)$ plane.
Lower panels)  The same as upper pannels but for $\Gamma/M=0.03$.
}
\label{fig:contour-scalar-gg}
\end{center}
\end{figure}

According to \eq{eq:twoSolutions}, 
there are actually two solutions for $\Gamma_{gg}/M$.
We update the previous analysis by considering  wider ranges of parameters
and including the contribution from photon fusion 
which is non-negligible when $\Gamma_{gg}/M$ is below $\mathcal{O}(10^{-4})$.  
In \fig{fig:contour-scalar-gg}, 
we show the regions of the partial decay widths allowed by 
the diphoton excess and GCE in wide ranges of parameter space in 
$(\Gamma_{gg}/M,\Gamma_{\gamma\gamma}/M)$ and 
$(\Gamma_{\chi\chi}/M,\Gamma_{\gamma\gamma}/M)$ planes.
The allowed regions are  at  $68\%$ and $95\%$ C.Ls. for two parameters, 
corresponding to $\Delta \chi^{2}=2.3$ and 6.0, respectively, 
together with the allowed regions by each individual experiment,
for the case of $\Gamma/M=0.06$ and 0.03.
It can be clearly seen that there is another solution located at
$\Gamma_{gg}/M\approx \Gamma/M$ which corresponds to the case where
the total width is dominated by gluon final states.
However, this solution is ruled out by the limit on the dijet production at Run-1,
as can be seen from the figure.

If the diphoton  events are generated from parton level $q\bar q$ annihilation,
the situation is  quite different.
For $q\bar q$ annihilation channel, 
the function $R_{q\bar q}$ is given by
\begin{align}
R_{q\bar{q}}=4\frac{\beta_{q}(4m_{\chi}^{2})}{\beta_{q}(M^{2})} \left(\frac{m_{\chi}}{M}\right)^{2}  ,
\end{align} 
Since $R_{q \bar q}$ is proportional to $(m_{\chi}/M)^{2}$ instead of 
$(m_{\chi}/M)^{4}$,
the lower limit on $\Gamma/M$ can be much smaller.
For DM annihilation dominantly into $b\bar b$ final states, 
\eq{eq:totalWidthLimit} can be rewritten as
\begin{align}
\left(\frac{\Gamma}{M} \right)_{\text{scalar},b\bar b}  
\gtrsim
\frac{\beta_{\chi}^{1/2}(M^{2})\langle \sigma v\rangle_{b\bar b}^{1/2} M }{4\pi^{1/2}}  .
\end{align}
Note that for $b\bar b$ final states,  it is determined by 
$\langle \sigma v\rangle_{b\bar b}$ alone,
as the  leading $m_{\chi}$ dependence cancels out in \eq{eq:totalWidthLimit}.
This observation holds for all the $q\bar q$ final states.
In the upper-right panel of \fig{fig:GCEfit}, 
we show the maximally allowed value of 
$\langle \sigma v\rangle$ as a function of DM particle mass
for three choices of $\Gamma/M=0.006$, 0.003 and 0.002, respectively.
If $\langle \sigma v\rangle_{b\bar b}$ is required  to be equal to $\langle \sigma v\rangle_{F}$, 
we find that the required $\Gamma/M$ should be above $\sim$ 0.006,
which is insensitive to the DM particle mass.

Using the best-fit values of $m_{\chi}$ and  
$\langle \sigma v\rangle_{b\bar b}$ for $b\bar b$ channel in \tab{tab:GCEdm}, 
the corresponding minimal value of the width-to-mass ratio is found to be 
\begin{align} 
\left(\frac{\Gamma}{M} \right)_{\text{scalar}, b\bar b}
\gtrsim 
3.6\times 10^{-3}
\left( \frac{M}{750~\text{GeV}}\right)
\left( \frac{\langle \sigma v\rangle_{b\bar b}}{1.4\times 10^{-26}~\mbox{cm}^{3}\mbox{s}^{-1}}\right)^{1/2}  .
\end{align} 
Similarly, for the $c\bar c$ channel, the minimal width is given by
\begin{align} 
\left(\frac{\Gamma}{M} \right)_{\text{scalar}, c\bar c}
\gtrsim 
3.0\times 10^{-3}
\left( \frac{M}{750~\text{GeV}}\right)
\left( \frac{\langle \sigma v\rangle_{c\bar c}}{0.95\times 10^{-26}~\mbox{cm}^{3}\mbox{s}^{-1}}\right)^{1/2}  .
\end{align} 
Thus for $q\bar q $ annihilation, the required minimal width-to-mass ratio
can be reduced to $\mathcal{O}(10^{-3})$, an order of magnitude lower
than that in the case of $gg$ fusion.

We perform  analogous  $\chi^{2}$ fits to 
the data of diphoton excess and the GCE in 
$b\bar b$ and $c\bar c$ channels
to determine the allowed values of the  the parameters 
$m_{\chi}$, $\Gamma_{\chi\chi}/M$, $\Gamma_{q\bar q}/M$ and 
$\Gamma_{\gamma\gamma}/M$ for two typical values of total width 
$\Gamma/M=0.06$ and 0.006, respectively.
For the diphoton excess, 
we take a naively weighted average of ATLAS and CMS results 
$\sigma_{\gamma\gamma}=8\pm2.1~\text{fb}$.
The Run-1 limits on dijet and diphoton productions are taken into account.
For the fit to the GCE, the data and the selection of the region of interest
in the sky are the same as the fit in Sec.~3. 
The results of the best-fit values and uncertainties of these parameters are 
summarized in \tab{tab:fitCombined}.
Compared with the fits to the GCE data alone, 
there are no significant changes in the determined DM particle mass.
The values of $\chi^{2}/\text{d.o.f}$ are also comparable, 
which indicates that the diphoton excess and GCE can be consistently 
explained in this model.

\begin{figure}[htb]
\begin{center}
\includegraphics[width=0.45\textwidth]{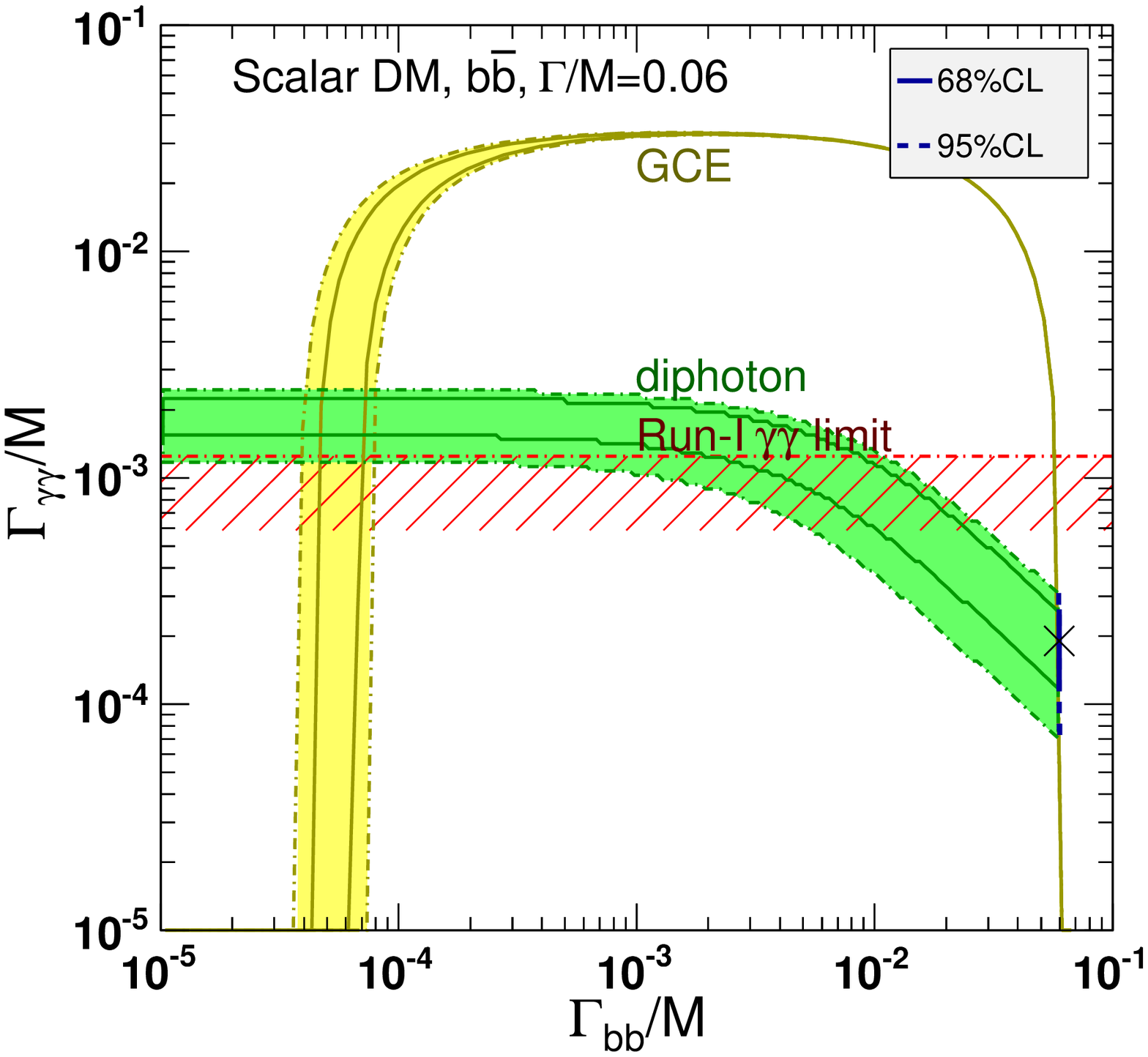}
\includegraphics[width=0.45\textwidth]{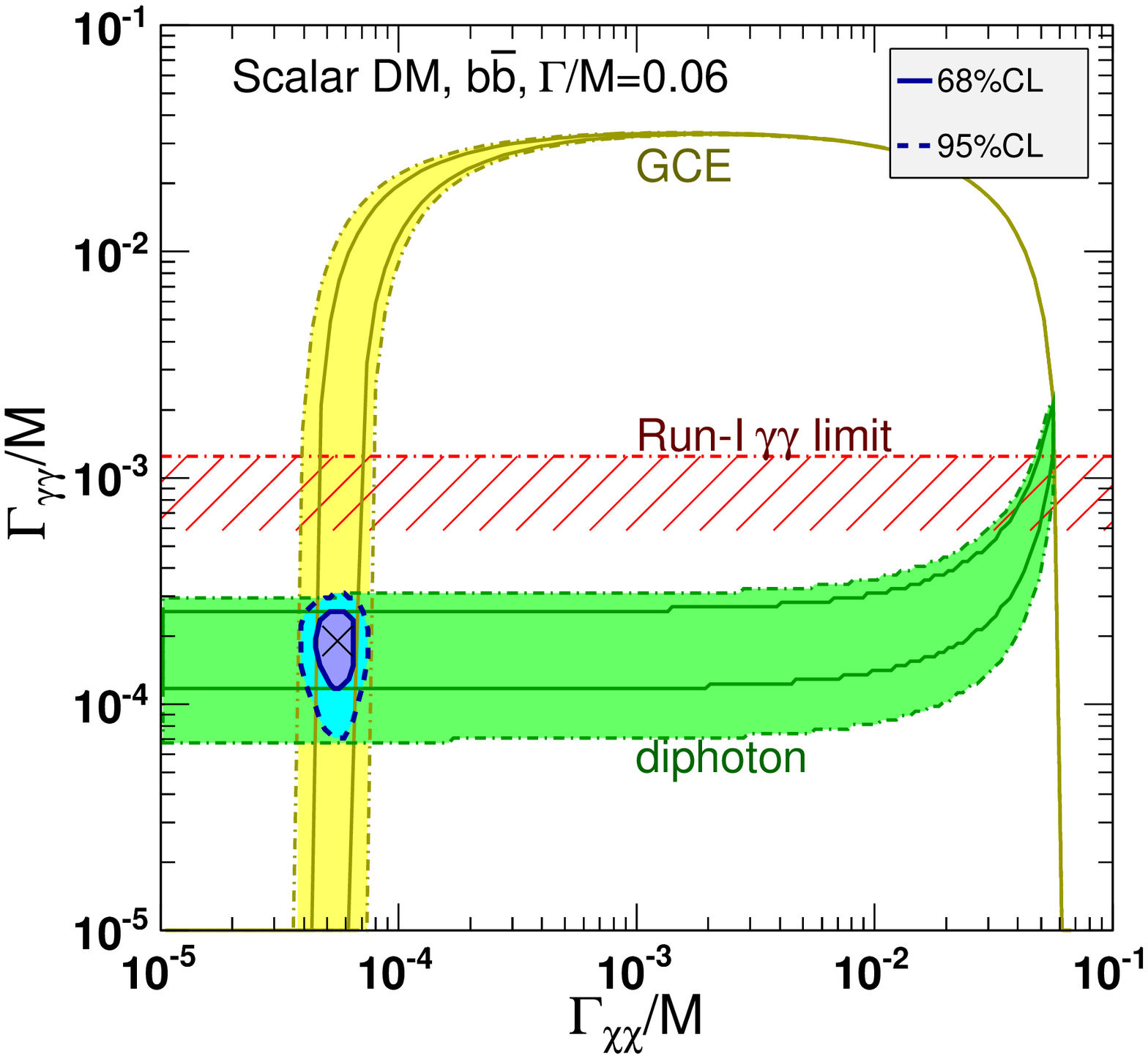}
\\
\includegraphics[width=0.45\textwidth]{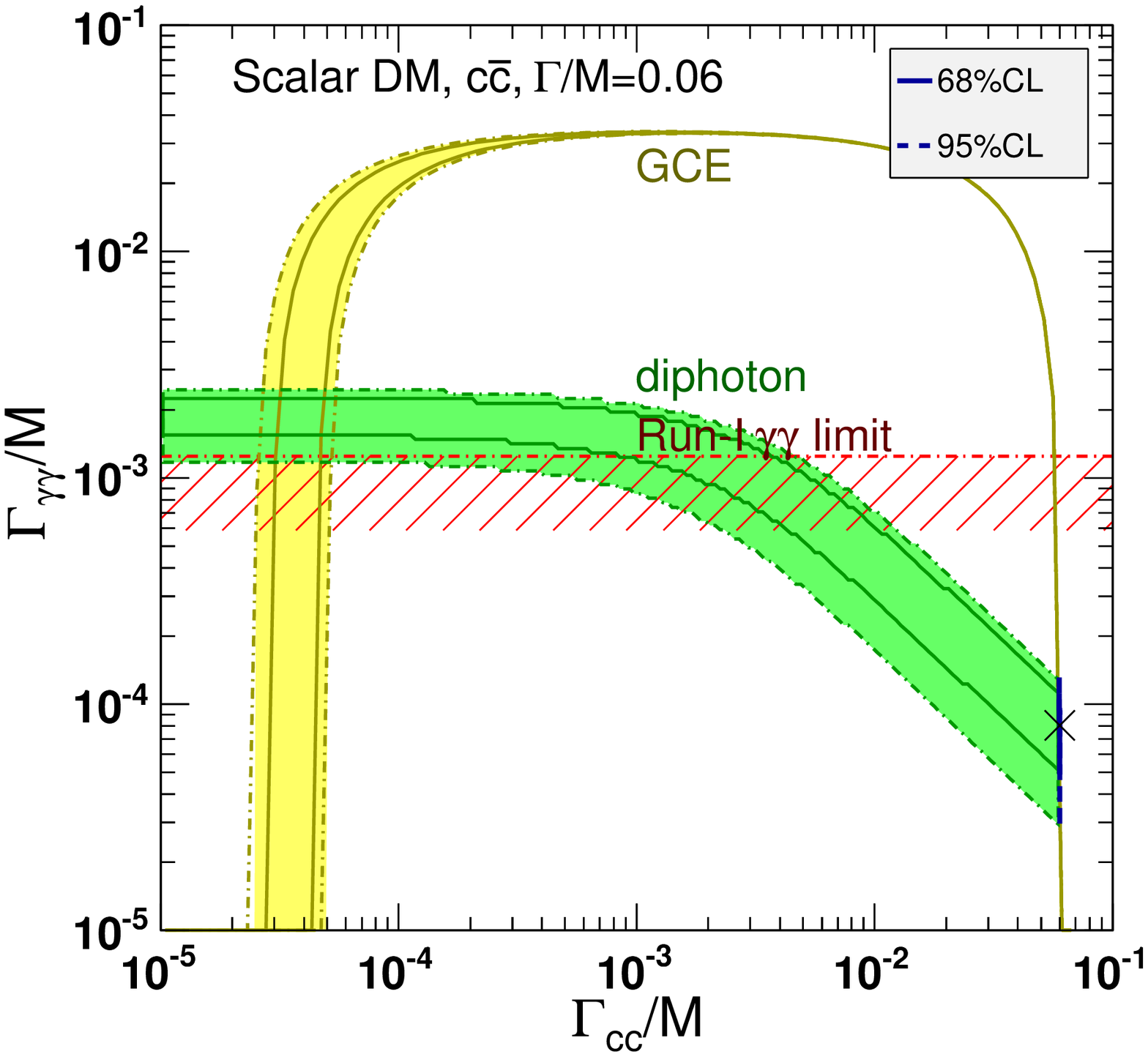}
\includegraphics[width=0.45\textwidth]{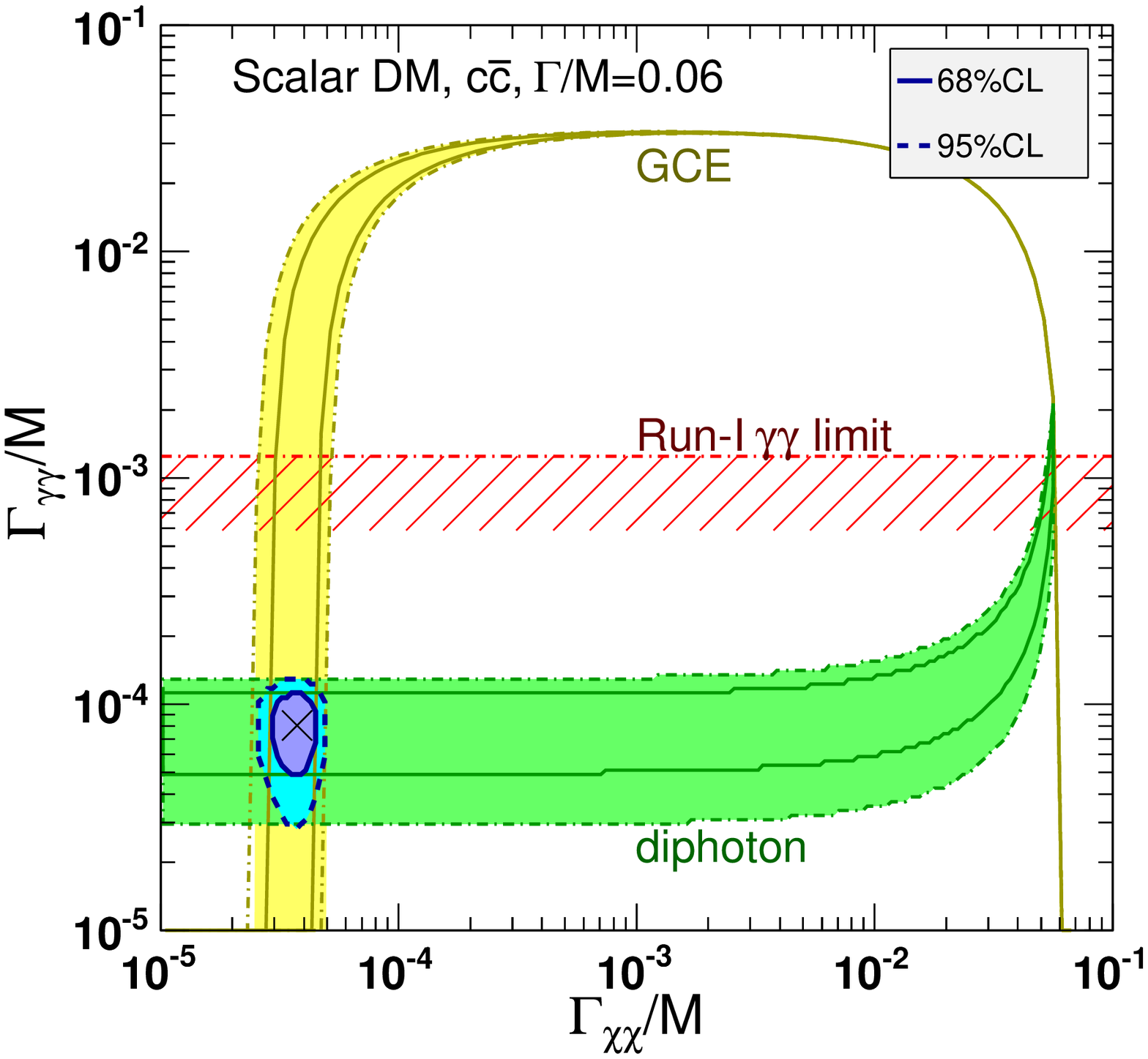}
\caption{
The same as \fig{fig:contour-scalar-gg}, but for scalar DM models with
$\phi$ generated from  $b\bar b$ (upper panels) and 
$c\bar c$ (lower panels) annihilation at the LHC for 
$\Gamma/M=0.06$.
}
\label{fig:contour-scalar-qq-0.06}
\end{center}
\end{figure}

\begin{figure}[htb]
\begin{center}
\includegraphics[width=0.45\textwidth]{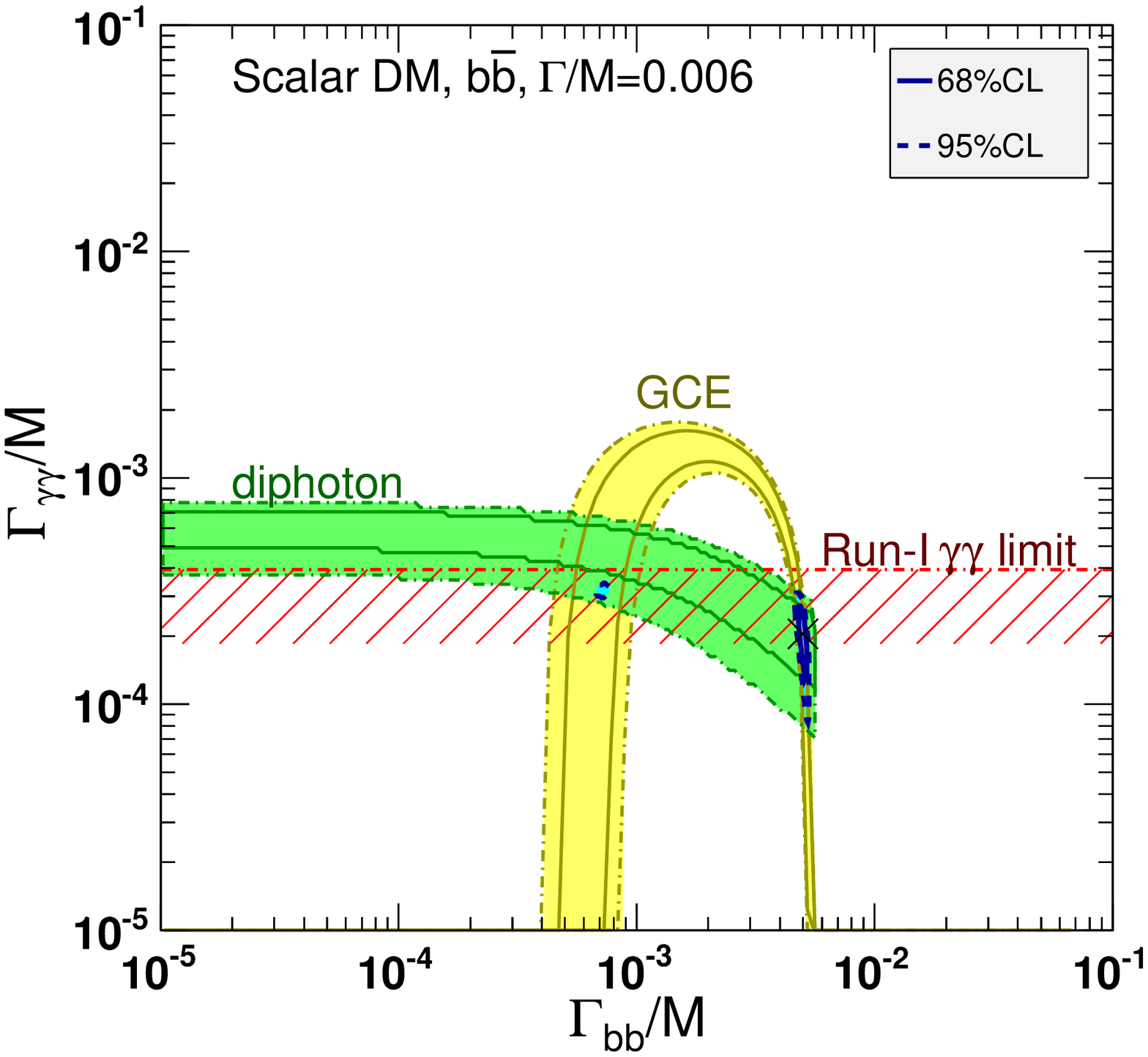}
\includegraphics[width=0.45\textwidth]{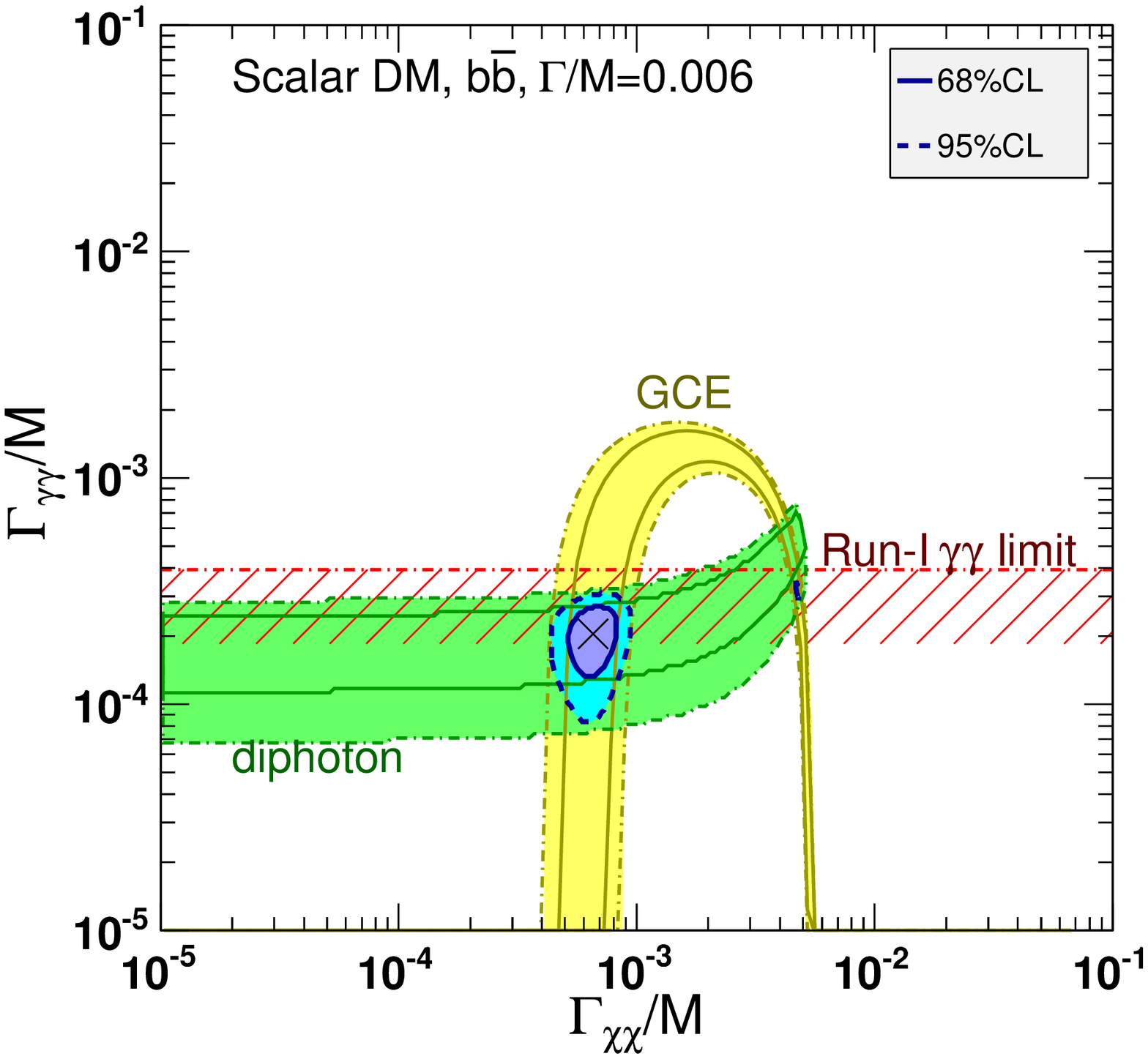}
\\
\includegraphics[width=0.45\textwidth]{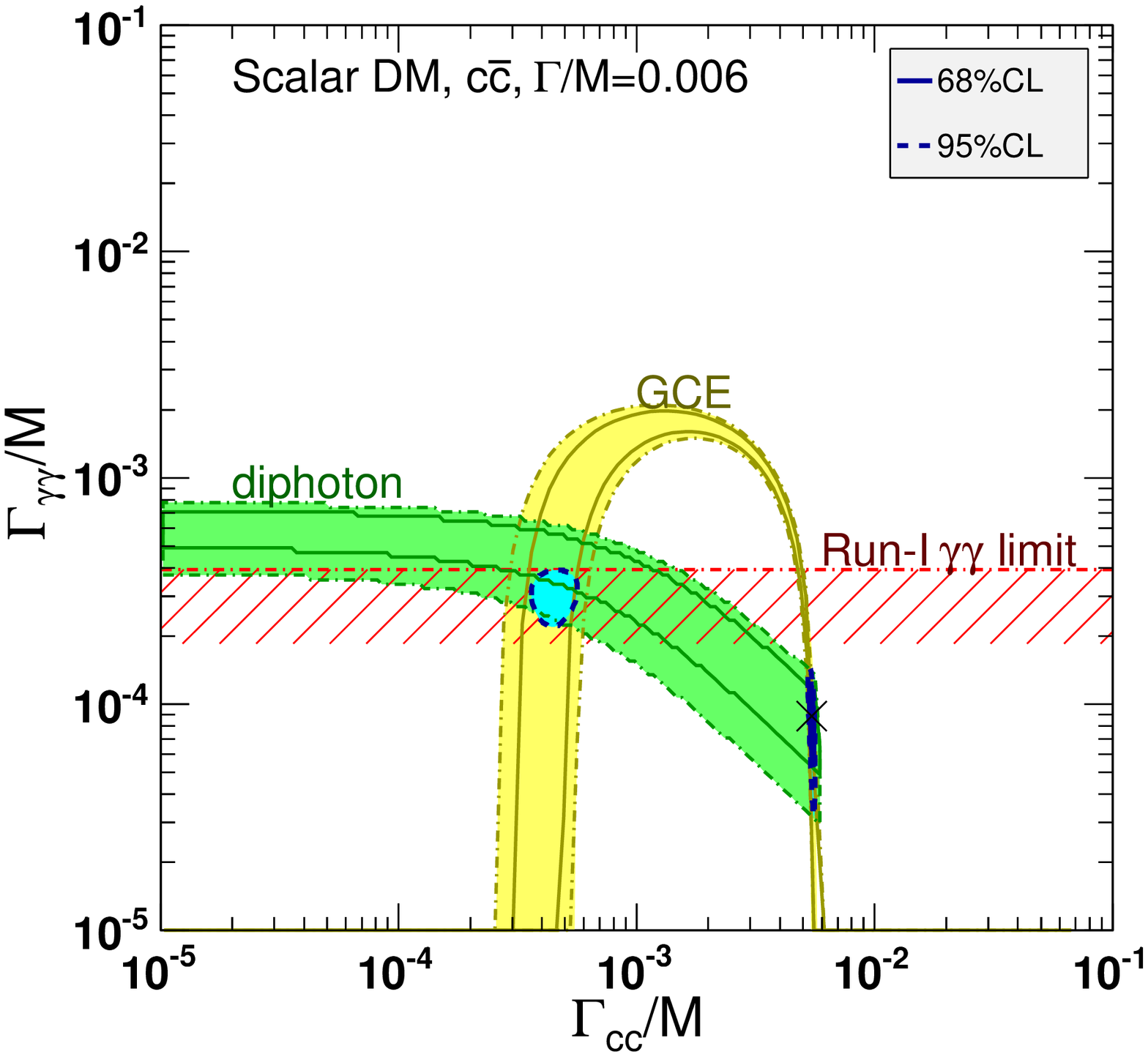}
\includegraphics[width=0.45\textwidth]{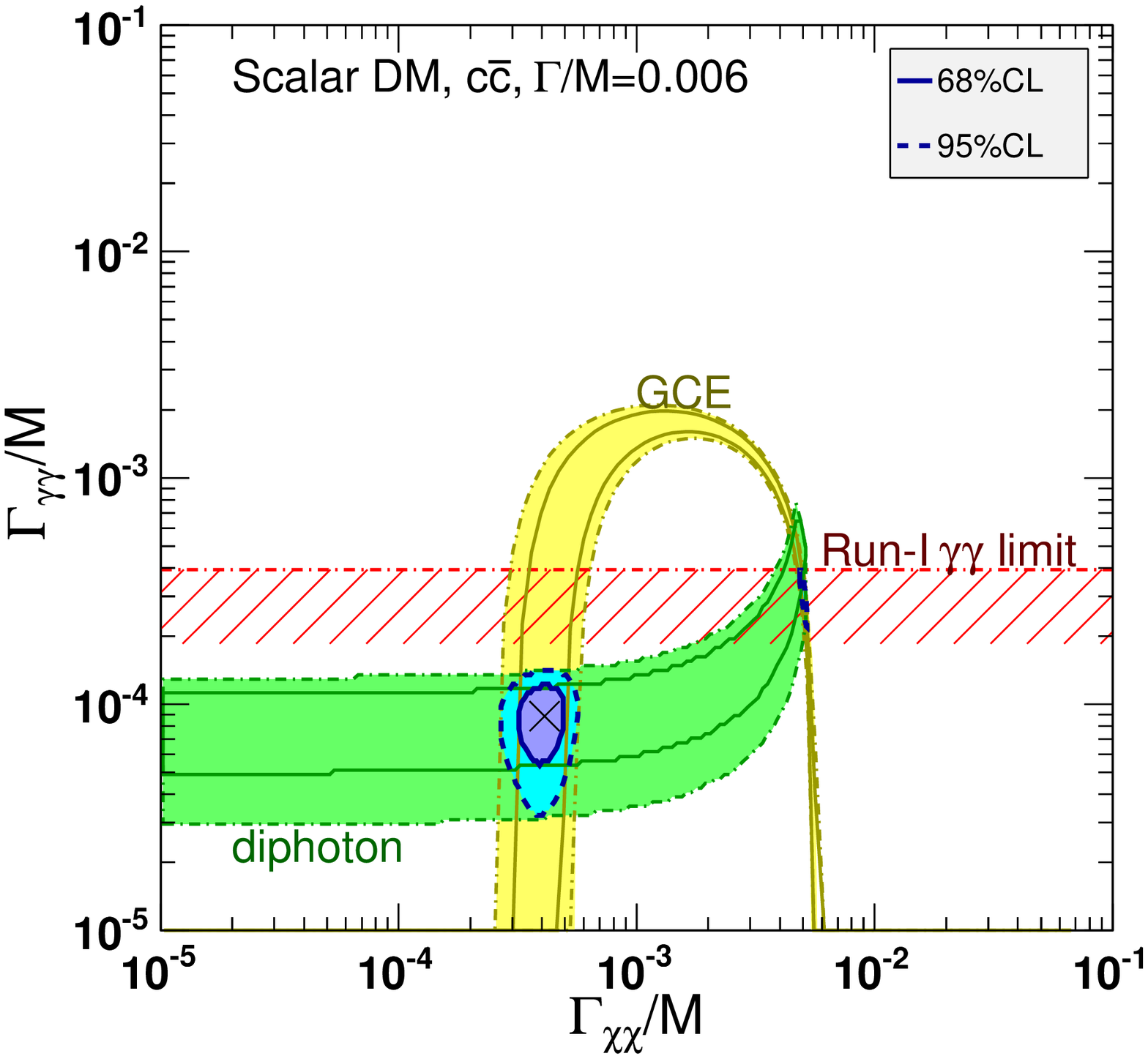}
\caption{
The same as \fig{fig:contour-scalar-qq-0.06}, but for $\Gamma/M=0.006$.
}
\label{fig:contour-scalar-qq-0.006}
\end{center}
\end{figure}

The allowed regions for the partial decay widths at  
$68\%$ and $95\%$ C.Ls.,
together with the allowed regions by each individual experiment,
for the case of $\Gamma/M=0.06\ (0.006)$ are shown in  \fig{fig:contour-scalar-qq-0.06} (\fig{fig:contour-scalar-qq-0.006}).
For the $q\bar q$ annihilation channels, 
the two solutions of \eq{eq:twoSolutions} can be  seen
as the two well-separated regions characterized by
\begin{align}
\frac{\Gamma_{\chi\chi}}{M} \approx  \frac{\Gamma}{M}, \quad
\frac{\Gamma_{q\bar q}}{M} \ll  \frac{\Gamma_{\chi\chi}}{M}  ,
& \qquad\qquad \text{(i)}  
\nonumber\\
\frac{\Gamma_{q\bar q}}{M} \approx  \frac{\Gamma}{M}, \quad
\frac{\Gamma_{\chi\chi}}{M} \ll  \frac{\Gamma_{q\bar q}}{M}  .
& \qquad\qquad \text{(ii)}  
\end{align}
The solution (i) corresponds to case of DM dominance while
the solution (ii) corresponds to the quark dominance in the total width.
In $q\bar q$ channels, the Run-1 dijet constraint does not apply.
However, the Run-1 constraint on the diphoton production cross section 
$\sigma_{\gamma\gamma}$ becomes relevant.  
In the large width case with $\Gamma/M=0.06$, 
for both the $b\bar b$ and $c\bar c$ channels, 
the solution (i) is ruled out by
the Run-1 limit on the diphoton production, 
as the required  $\Gamma_{\gamma\gamma}$ is above $\mathcal{O}(10^{-3})$.
The solution (ii) is consistent with the data, and 
the favoured $\Gamma_{\chi\chi}/M$ are of $\mathcal{O}(10^{-5})$, and 
$\Gamma_{\gamma\gamma}/M$ are of $\mathcal{O}(10^{-4})$.
In the solution (ii), from $\Gamma_{q\bar q}/M\approx \Gamma/M=0.06$,
the size of the Yukawa coupling is found to be $y_{q}^{2}\approx 0.5$
which is marginally within the perturbative regime.
In the small width case with $\Gamma/M=0.006$, 
for  $b\bar b$ channel, both the solutions are close to 
the Run-1 diphoton limit. 
But the solution (ii) is favoured against solution (i).
For the  $c\bar c$ channel, the situation is similar.
In the small width case $\Gamma/M=0.006$,
the favoured $\Gamma_{\chi\chi}/M$ is comparable with 
$\Gamma_{\gamma\gamma}$, both are of  $\mathcal{O}(10^{-4})$.
The determined values of $m_{\chi}$ and 
$\langle \sigma v\rangle_{b\bar b, c\bar c}$ for the solution (ii)
are listed in \tab{tab:fitCombined}.

\begin{figure}[htbp]
\begin{center}
\includegraphics[width=0.45\textwidth]{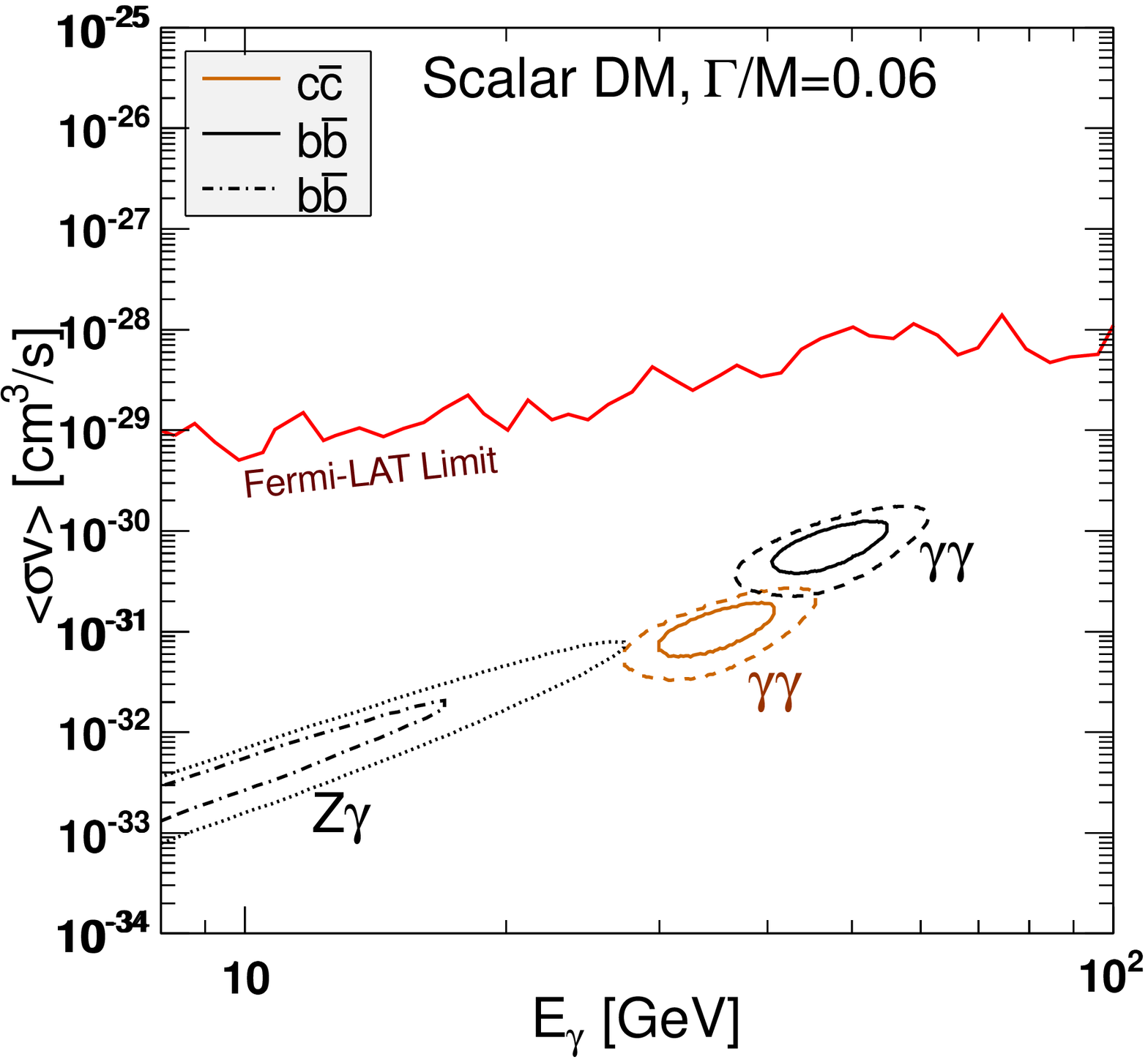}
\includegraphics[width=0.45\textwidth]{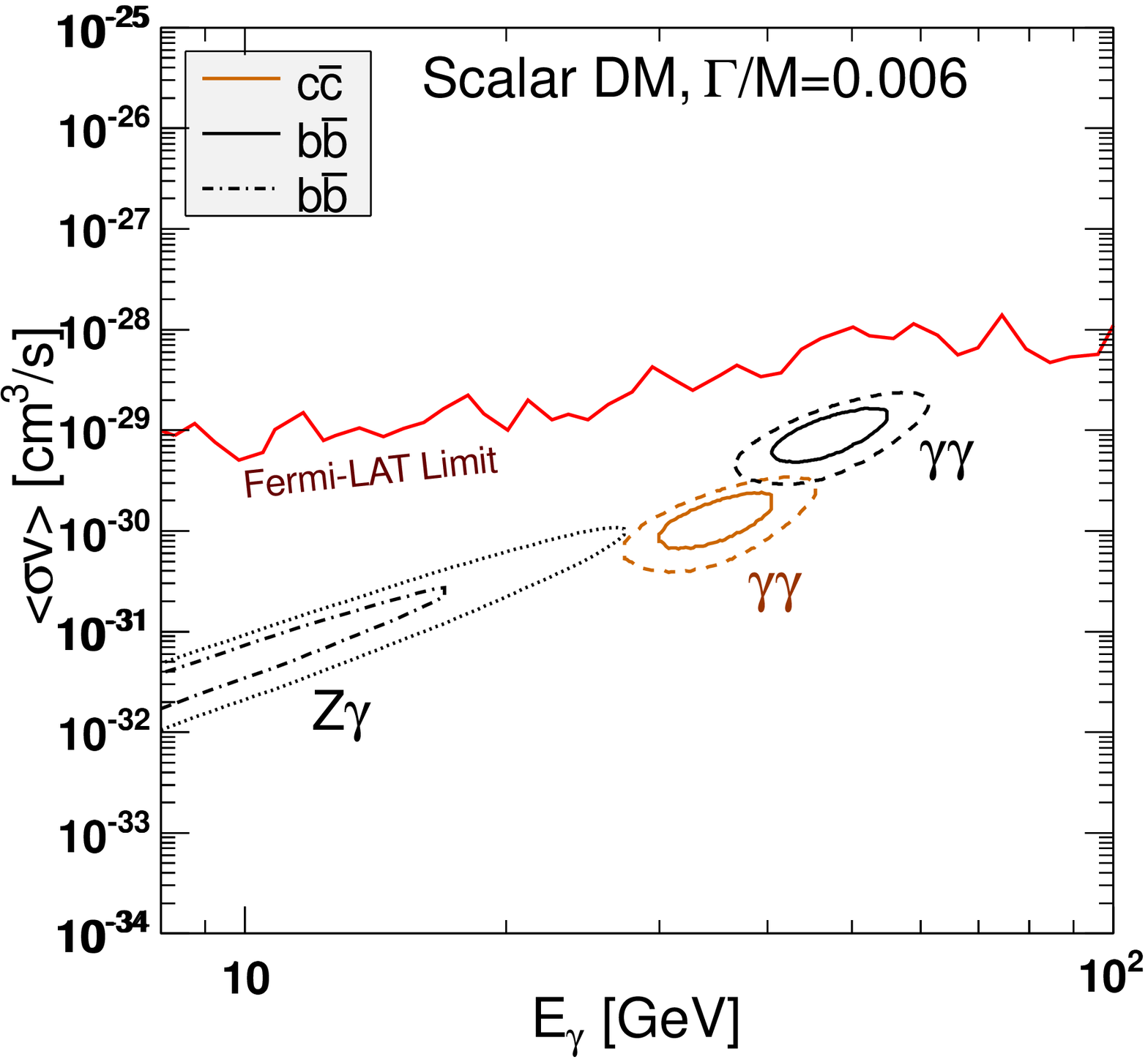}
\caption{
Left)
Predictions for $\langle \sigma v \rangle_{\gamma\gamma}$ and 
$\langle \sigma v \rangle_{Z\gamma}$ as a function of photon energy
in the scalar DM model with $\phi$ coupling dominantly with $b\bar b$ and $c\bar c$,
using the allowed  parameters from the fit to the data of diphoton exces and GCE, 
for $\Gamma/M=0.06$. The exclusion limits at $95\%$ C.L. of Fermi-LAT 
\cite{Ackermann:2015lka} for region R16 are also shown.}
\label{fig:gammaline-scalar-qq}
\end{center}
\end{figure}

Since in the $q\bar q$ channel, the total width is not DM dominated.
The predicted cross section for gamma-ray lines which is proportional to 
$\Gamma_{\chi\chi}\Gamma_{\gamma\gamma}$
 can be smaller. 
In \fig{fig:gammaline-scalar-qq}, we give the predicted cross sections for DM annihilation into $\gamma\gamma$ which gives rise to the gamma-ray spectral lines,
based on the parameters determined from the fit results listed in \tab{tab:fitCombined}
for $b\bar b$ and $c\bar c$ channels with 
two different values of $\Gamma/M=0.06$ and 0.006, respectively.
In all the cases  the predictions are well below the current upper limits set by Fermi-LAT.
For $\Gamma/M=0.06$, the predicted  cross section 
$\langle \sigma v \rangle_{\gamma\gamma}$ is 
$\sim 5\times 10^{-31}~\text{cm}^{3}\text{s}^{-1}$ for $b\bar b$ channel,
and $\langle \sigma v \rangle_{Z\gamma}$ is below
$\sim 10^{-31}~\text{cm}^{3}\text{s}^{-1}$.
For $c\bar c$ channel, the predicted  cross section 
$\langle \sigma v \rangle_{\gamma\gamma}$ is 
$\sim 1\times 10^{-31}~\text{cm}^{3}\text{s}^{-1}$.
The $Z\gamma$ final state is kinematically forbidden  due to 
th low mass of the DM particle.
For $\Gamma/M=0.06$, the predictions are relatively higher,
which is due to the fact that a larger $\Gamma_{\chi\chi}/M$ of $\mathcal{O}(10^{-4})$
is favoured.

\begin{table}\begin{center}
   \begin{tabular}{llccccc}
    \hline\hline
    Channel & $\Gamma/M$ & $m_\chi$ (GeV)  & $\Gamma_{\gamma\gamma}/M (\times 10^{-4}) $  & $\Gamma_{\chi\chi}/M $ & $\chi^2_{\text{min}}/\text{d.o.f.} $ & $p$-value \\
    \hline
	Scalar DM,  $b\bar{b}$ & 0.06 & $46.15^{+5.81}_{-3.53}$ & $1.90^{+0.49}_{-0.50}$ & $5.52^{+0.68}_{-0.68}\times10^{-5}$ & 25.418/23 & 0.33 \\
	 & 0.006 & $46.15^{+5.81}_{-3.53}$ & $2.04^{+0.46}_{-0.48}$ & $6.56^{+0.97}_{-0.92}\times10^{-4}$ & 24.578/23 & 0.37 \\ %
    Scalar DM, $c\bar{c}$ & 0.06 & $35.54^{+3.10}_{-4.12}$ & $0.81^{+0.21}_{-0.21}$ & $3.80^{+0.47}_{-0.47}\times10^{-5}$ & 26.664/23 & 0.27 \\ %
   & 0.006 & $35.54^{+3.10}_{-4.12}$ & $0.89^{+0.23}_{-0.23}$ & $4.11^{+0.56}_{-0.54}\times10^{-4}$ & 26.311/23 & 0.29 \\ %
   \hline
	Fermionic DM, $b\bar{b}$ & 0.06 & $46.20^{+6.37}_{-2.68}$ & $4.55^{+2.39}_{-2.11}$ & $ 3.49^{+0.85}_{-1.75}\times10^{-2}$ & 24.906/23 & 0.36 \\ %
 Fermionic DM,  $c\bar{c}$ & 0.06 & $36.48^{+3.13}_{-2.11}$ & $1.62^{+0.82}_{-0.55}$ & $3.01^{+0.81}_{-0.85}\times10^{-2}$ & 27.048/23 & 0.25 \\ %
    \hline\hline
    \end{tabular}%
\caption{
Values of DM mass $m_{\chi}$,  partial width-to-mass ratios 
$\Gamma_{\gamma\gamma}/M$ and $\Gamma_{\chi\chi}/M$
determined from
combined fits to both the LHC diphoton excess and the GCE 
in scalar and fermoinic DM models with constraints from Run-1 data
on the dijet and diphoton searches included.
For scalar DM models, 
the results for the cases of  $\phi$ coupling dominantly to 
$b\bar b$ or  $c\bar c$ with total width $\Gamma/M=0.06$ and 0.006 are given. 
For fermionic DM models, the results are for $\Gamma/M=0.06$.
The corresponding $\chi^{2}/\text{d.o.f}$ and $p$-values for each fit are also shown. 
}
\label{tab:fitCombined}
\end{center}\end{table}

\subsection{Fermionic DM} 
For fermionic DM, 
we shall focus on the case where $\chi$ is a Majorana fermion.  
The results for the Dirac  DM particle can be obtained 
in a straight forward way. 
The Lagrangian for the Majorana DM particle and 
its interaction with $\phi$ is given by
\begin{align}
\mathcal{L} & \supset
\frac{1}{2}\bar\chi (i\gamma^{\mu}\partial_{\mu}-m_{\chi})\chi
-\frac{1}{2} y_{\chi}\bar\chi i\gamma^{5} \chi \phi  .
\end{align}
For the DM annihilation into $gg$ and $q\bar q$ final states, 
the corresponding $R_{X\bar{X}}$ factors are
\begin{align}
R_{gg}=64 \left( \frac{m_{\chi}}{M}\right)^{6}
\quad \text{and} \quad
R_{q\bar{q}}=16\frac{\beta_{q}(4m_{\chi}^{2})}{\beta_{q}(M^{2})}
\left( \frac{m_{\chi}}{M}\right)^{4}  .
\end{align}
For $gg$ annihilation final states in this model, 
the \eq{eq:totalWidthLimit} can be written as
\begin{align}
\left(\frac{\Gamma}{M} \right)_{\text{fermion},gg}  
\geq
\frac{\beta_{\chi}^{1/2}(M^{2})\langle \sigma v\rangle_{gg}^{1/2}M^{3}}{8\pi^{1/2}m_{\chi}^{2}}  .
\end{align}
Since in this model the $R_{gg}$ factor is proportional to $(m_{\chi}/M)^{6}$,
the required total width is quite large.
In the lower-left panel of \fig{fig:GCEfit}, 
we show the upper limit on $\langle \sigma v\rangle_{gg}$ as a function of $m_{\chi}$
for three choices of $\Gamma/M=0.5$, 0.2 and 0.06, respectively.
For DM particle mass below $\sim100$ GeV, 
the value of $\langle \sigma v\rangle_{gg}$ is 
far below the typical thermal cross section. 
For a consistent explanation to the DM relic density, 
the required DM particle mass should be above $\sim 150$ GeV, 
for $\Gamma/M=0.06$. 
In fermionic DM model, 
the factor $R_{q\bar q}$ is the same as $R_{gg}$ in the scalar DM model.
Thus the upper limit on the  cross sections can be obtained from 
that in the scalar DM model by a rescaling factor $1/s_{\chi}=1/4$.

For $gg$ channel,
using the best-fit values of $m_{\chi}$ and  
$\langle \sigma v\rangle_{gg}$  in \tab{tab:GCEdm}, 
the corresponding minimal value of the width-to-mass ratio is found to be
\begin{align} 
\left(\frac{\Gamma}{M} \right)_{\text{fermion}, gg}
\gtrsim 
0.31
\left( \frac{M}{750~\text{GeV}}\right)^{3}
\left( \frac{62~\text{GeV}}{m_{\chi}}\right)^{2}
\left( \frac{\langle \sigma v\rangle_{gg}}{1.96\times 10^{-26}~\mbox{cm}^{3}\mbox{s}^{-1}}\right)^{1/2}  .
\end{align} 
Such a large with is not favoured by the 
current experimental data  and is theoretically unnatural.

\begin{figure}[htb]
\begin{center}
\includegraphics[width=0.45\textwidth]{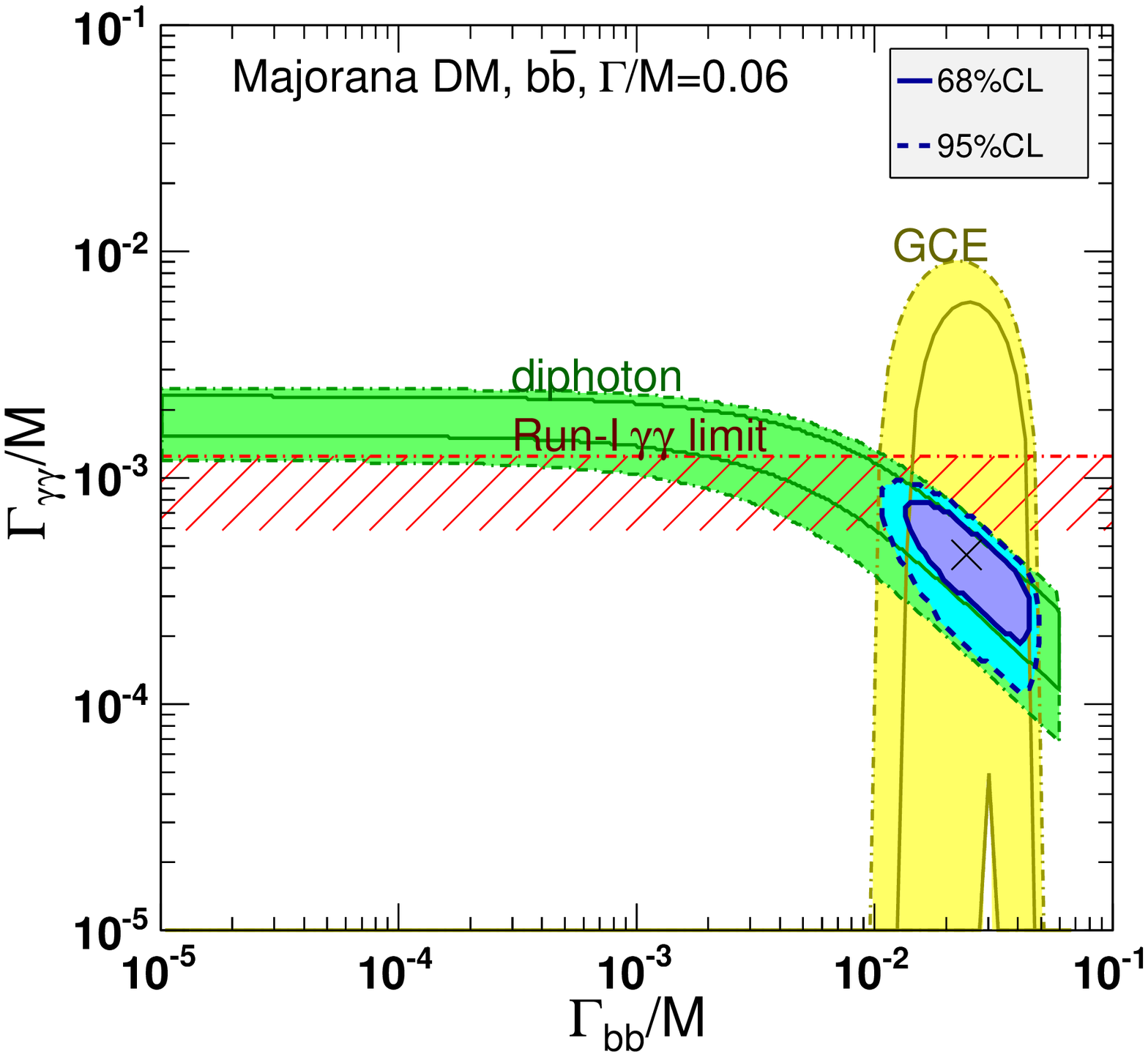}
\includegraphics[width=0.45\textwidth]{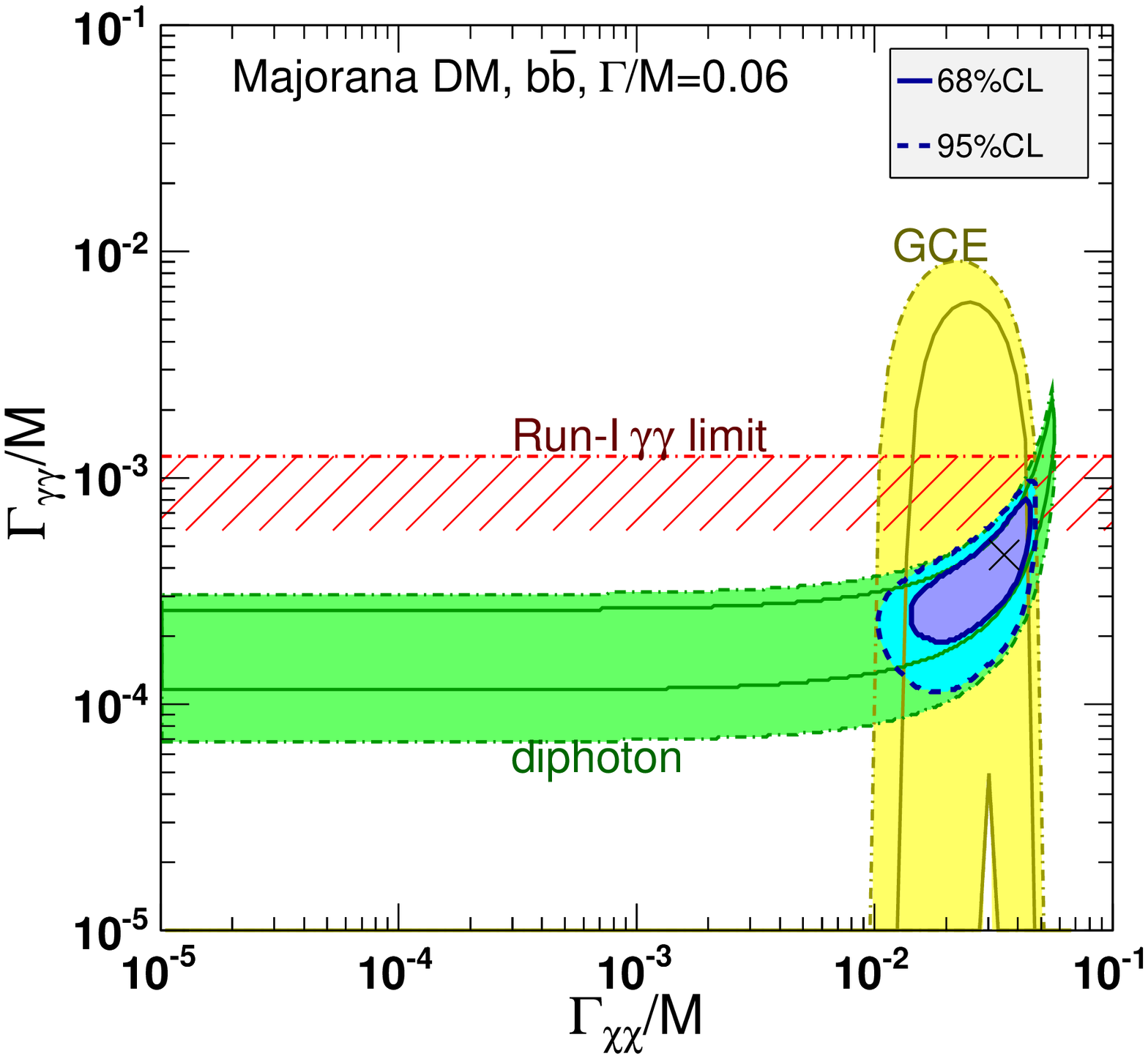}
\\
\includegraphics[width=0.45\textwidth]{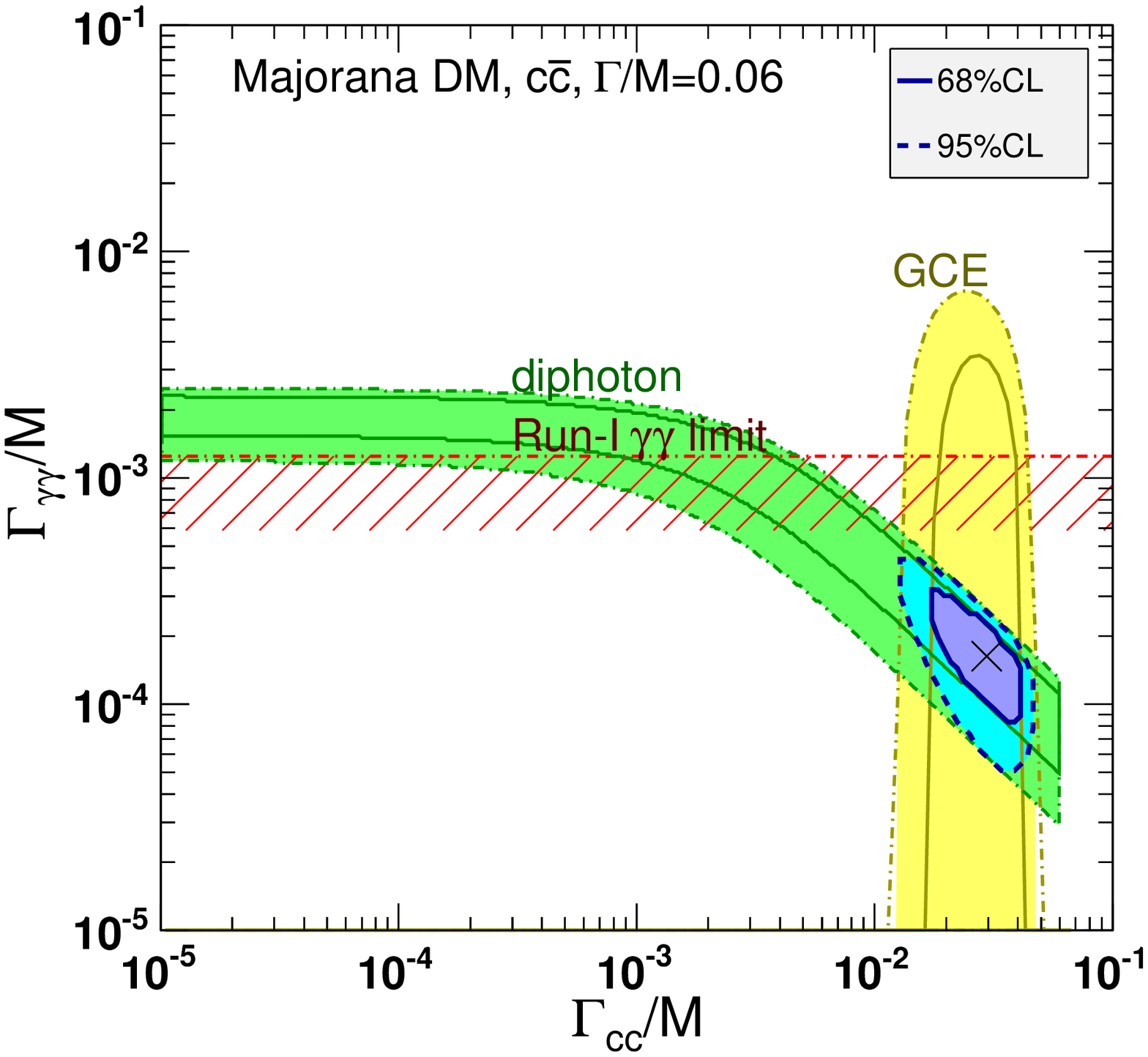}
\includegraphics[width=0.45\textwidth]{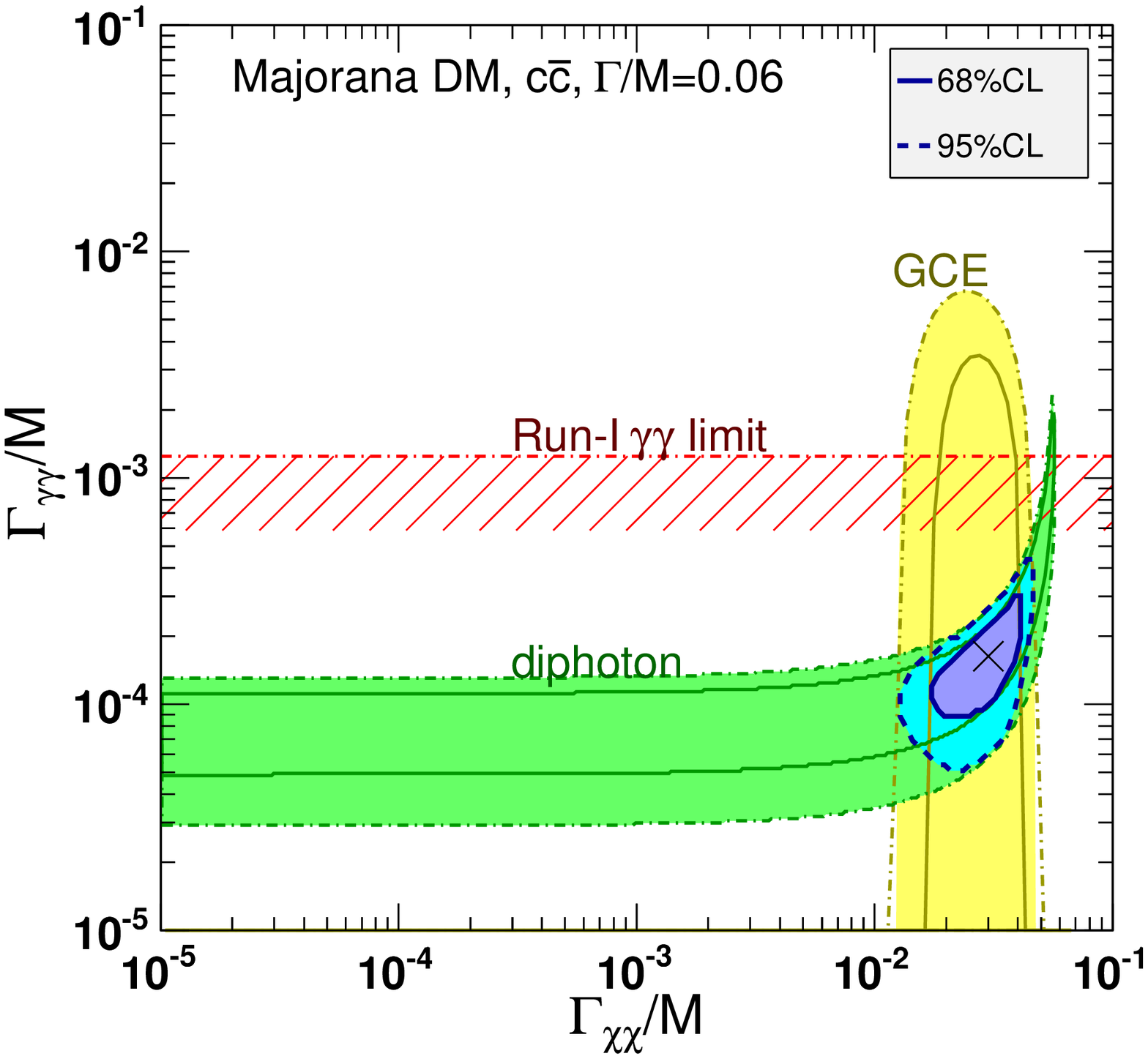}
\caption{
The same as \fig{fig:contour-scalar-qq-0.06} but for 
the Majoranna fermionic DM model.
}
\label{fig:contour-Majorana-qq-0.06}
\end{center}
\end{figure}

For $q\bar q$-channel, since $R_{q\bar q}$ is proportional to $(m_{\chi}/M)^{4}$, 
the required total width is similar to the case of $gg$-channel of scalar DM.
For $b\bar b$ channel it is found that
\begin{align} 
\left(\frac{\Gamma}{M} \right)_{\text{fermion}, b\bar b}
\gtrsim 
0.058
\left( \frac{M}{750~\text{GeV}}\right)^{2}
\left( \frac{46~\text{GeV}}{m_{\chi}}\right)
\left( \frac{\langle \sigma v\rangle_{gg}}{1.42\times 10^{-26}~\mbox{cm}^{3}\mbox{s}^{-1}}\right)^{1/2} ,
\end{align} 
and the result is similar for the $c\bar c$ channel
\begin{align} 
\left(\frac{\Gamma}{M} \right)_{\text{fermion}, c\bar c}
\gtrsim 
0.062
\left( \frac{M}{750~\text{GeV}}\right)^{2}
\left( \frac{35.5~\text{GeV}}{m_{\chi}}\right)
\left( \frac{\langle \sigma v\rangle_{gg}}{0.95\times 10^{-26}~\mbox{cm}^{3}\mbox{s}^{-1}}\right)^{1/2}  .
\end{align} 
We perform $\chi^{2}$-fit to the diphoton and GCE data in 
the fermionic DM model for $b\bar b$ and $c\bar c$  channels with 
$\Gamma/M=0.06$.
The determined parameters are shown in \tab{tab:fitCombined}, and
the allowed regions of the parameters in 
$(\Gamma_{b\bar b}/M, \Gamma_{\gamma\gamma}/M)$ and  
$(\Gamma_{\chi\chi}/M, \Gamma_{\gamma\gamma}/M)$ planes
are shown in \fig{fig:contour-Majorana-qq-0.06}. 
Compared with the same channel in the scalar DM model, 
a visible change in the allowed regions is that 
the regions corresponding to the  two solutions merge together,
which is due to the fact that in fermionic DM models, 
the value of $\Delta$ is quite small as 
the minimally required width is close to 0.06.
The determined values of $\Gamma_{q\bar q}/M$ and $\Gamma_{\chi\chi}/M$
are roughly the same order of magnitude about $\mathcal{O}(10^{-2})$.
The allowed regions are consistent with 
the Run-1 limit on cross section of the diphoton production .

\begin{figure}[htbp]
\begin{center}
\includegraphics[width=0.45\textwidth]{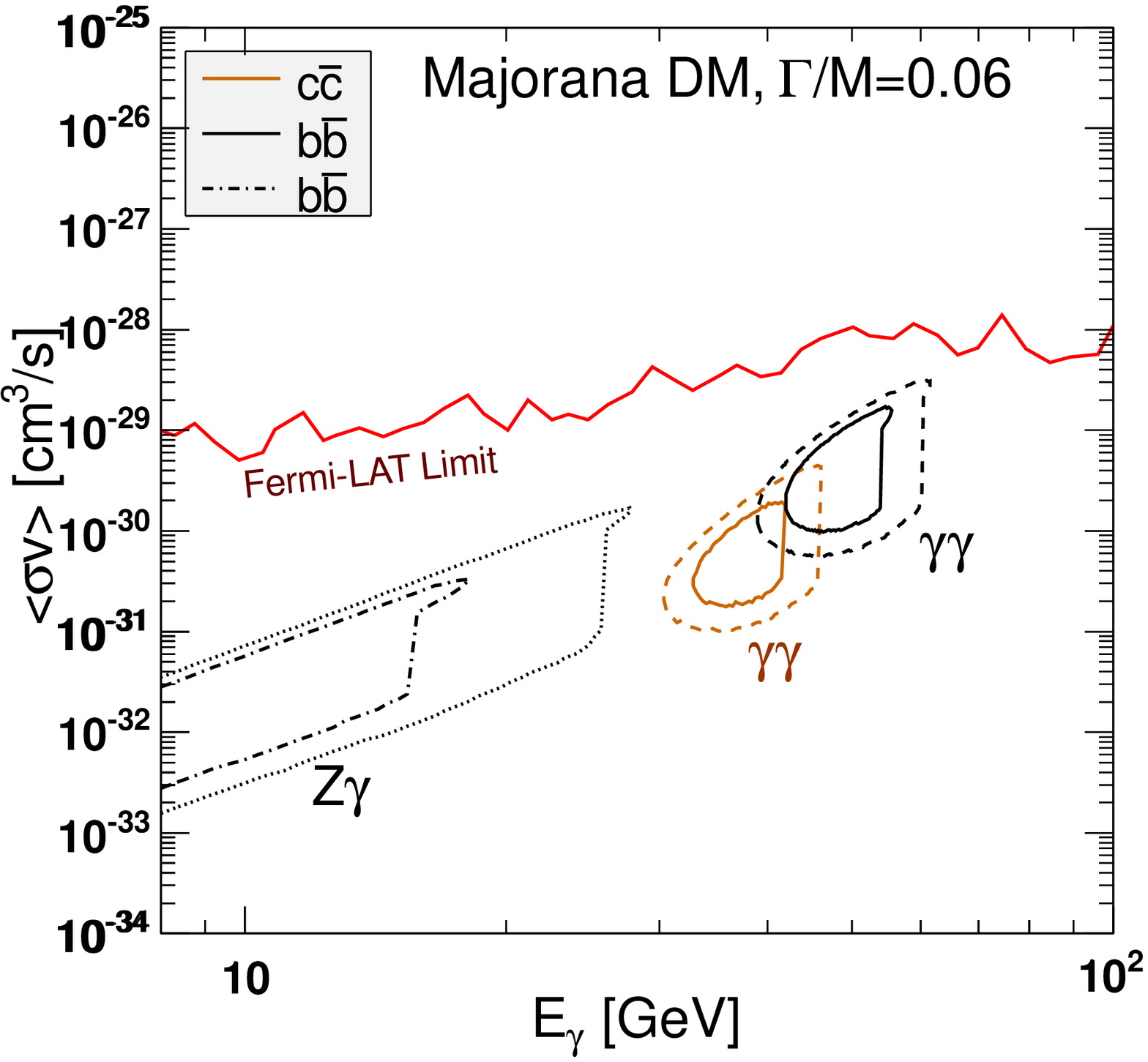}
\caption{
The same as \fig{fig:gammaline-scalar-qq}, but for Majorana DM model and 
$\Gamma/M=0.06$.
}
\label{fig:gammaline-fermionic-qq}
\end{center}
\end{figure}

Since in the fermionic DM model, 
$\Gamma_{\chi\chi}/M$ can reach $\mathcal{O}(10^{-2})$, 
it is expected that the predicted cross sections for the gamma-ray line
are significantly larger than that in the scalar DM model.
In \fig{fig:gammaline-fermionic-qq}, 
we show the predicted cross sections in this model for $b\bar b$ and 
$c\bar c$ channel with $\Gamma/M=0.06$.
The cross section can reach $\mathcal{O}(10^{-29}) \text{ cm}^{3}\text{s}^{-1}$,
which is very close to the current Fermi-LAT limit,  and
can be tested in the future by Fermi-LAT, HESS and CTA.

In the case where $\chi$ is Dirac, 
the corresponding values of $R_{X\bar X}$ are the same. 
However, the required product of $(\Gamma_{\chi\bar\chi}/M)(\Gamma_{X\bar X})/M$
increases by a factor of four from Eqs. (\ref{eq:sigmav}) and (\ref{eq:fluxAvg}). 
Thus the required total width $\Gamma/M$ is expected to be larger compared 
with all the cases of Majorana DM.

\subsection{Vector DM} 
In the case where the DM particle is a Majorana fermion, 
the Lagrangian for DM and its interaction with $\phi$ is given by
\begin{align}
\mathcal{L} & \supset
\frac{1}{4 } \chi_{\mu\nu}\chi^{\mu\nu}
-\frac{1}{2} m_{\chi}^{2} \chi_{\mu}\chi^{\mu}
-\frac{1}{2} g_{\chi} \phi \chi_{\mu}\chi^{\mu}.
\end{align}
For $\text{\ensuremath{\chi\chi\to\phi\to}}gg, q\bar{q}$, 
the corresponding $R_{X\bar{X}}$ factors are given by 
\begin{align}
R_{gg}=\frac{192m_{\chi}^{8}}{M^{8}T(m_{\chi}/M)}, \quad
R_{q\bar{q}}=\frac{64m_{\chi}^{6}\beta_{q}(4m_{\chi}^{2})}{M^{6}\beta_{q}(M^{2})T(m_{\chi}/M)} ,
\end{align}
where $T(x)=1-4x^{2}+12x^{4}$.
Since in vector DM model $R_{gg}\propto (m_{\chi}/M)^{8}$ and
$R_{q\bar q}\propto (m_{\chi}/M)^{6}$,
it is expected that a very large $\Gamma/M$ is required. 
For $gg$ channel, the minimally required width is given by 
\begin{align} 
\left(\frac{\Gamma}{M} \right)_{\text{vector}, gg}
\gtrsim 
3.3
\left( \frac{M}{750~\text{GeV}}\right)^{4}
\left( \frac{62~\text{GeV}}{m_{\chi}}\right)^{3}
\left( \frac{\langle \sigma v\rangle_{gg}}{1.96\times 10^{-26}~\mbox{cm}^{3}\mbox{s}^{-1}}\right)^{1/2},
\end{align} 
and for $q\bar q$ channel
\begin{align} 
\left(\frac{\Gamma}{M} \right)_{\text{fermion}, b\bar b}
\gtrsim 
0.73
\left( \frac{M}{750~\text{GeV}}\right)^{3}
\left( \frac{46~\text{GeV}}{m_{\chi}}\right)^{2}
\left( \frac{\langle \sigma v\rangle_{gg}}{1.42\times 10^{-26}~\mbox{cm}^{3}\mbox{s}^{-1}}\right)^{1/2} .
\end{align} 
The results for the $c\bar c$ channel is similar to that in the $b\bar b$ channel.
Thus in all the cases, the required $\phi$ width are too large and 
already ruled out by the current Run-2 data, 
which indicates that the vector DM model can not provide a 
consistent explanation to the LHC Run-2 diphoton excess and the GCE.

\section{Conclusions}

In summary, we have investigate the  conditions for 
a consistent explanation for  possible the LHC diphoton excess and GCE,
especially the requirement on total width of $\phi$ in 
a wide range of DM models where
the DM particle can  be  scalar, fermionionic and vector, and
$\phi$ can be generated  by 
$s$-channel gluon fusion  or quark-antiquark annihilation ($b\bar b $ and $c\bar c$) 
at parton level.
We have shown that the required $\Gamma/M$ is determined by
a single parameter  proportional to $(m_{\chi}/M)^{n}$.
We have found that three models can explain the two excesses successfully:
i) scalar DM model with $\phi$ coupling dominantly with $q\bar q$.
the minimally required  $\Gamma/M$ can be as low as $\mathcal{O}(10^{-3})$;
ii)  scalar DM model with $\phi$ coupling dominantly with $gg$, the required  $\Gamma/M$ is about $\mathcal{O}(10^{-2})$;
iii) fermionic DM model with coupling dominantly with $q\bar q$, the required  $\Gamma/M$ reaches $\mathcal{O}(10^{-2})$.
Other models such as the vector DM model requires larger $\Gamma/M$ of order one
which is already disfavoured by the current data.
For the same DM model, the required width of $\phi$ is always smaller
in $q\bar q$ channel than that in the $gg$ channel.
For the DM models which can simultaneously account for the diphoton excess and the GCE, 
the predicted cross sections for gamma-ray line are typically of 
$\mathcal{O}(10^{-30})~\text{cm}^{3}\text{s}^{-1}$, 
which is close to  the current limits imposed by the  Fermi-LAT data.
These models can be distinguish soon by the updated LHC data, 
through the measurement of the total width,
and Fermi-LAT data on the gamma-ray line searches in the near future.

{\bf Acknowledgements.}
This work is supported 
by
the NSFC  under Grants
No. 11335012 and
No. 11475237.

\bibliographystyle{JHEP} %
\bibliography{diphotonGC_II,2hdm750,leftRight750,misc}
\end{document}